\documentclass[journal]{IEEEtran}
\usepackage{graphics} 
\usepackage{epsfig} 
\usepackage{times} 
\usepackage{amsmath} 
\usepackage{amssymb}  
\usepackage{cite}
\usepackage{epstopdf}
\usepackage{subfigure,color,balance}
\usepackage{verbatim}
\usepackage{cases}
\usepackage{enumerate}
\usepackage{enumitem}
\usepackage{booktabs}
\usepackage{balance}
\usepackage{mathbbol}
\usepackage{algorithm}
\usepackage{algorithmicx}
\usepackage{algpseudocode}
\usepackage{mathrsfs}
\usepackage{threeparttable}

\newcommand{\method}[1]{\texttt{#1}}

\usepackage[colorlinks=true,      
linkcolor=black,      
citecolor=black,      
filecolor=black,      
urlcolor=blue]{hyperref}

\newtheorem{definition}{Definition}

\newtheorem{proposition}{Proposition}
\newtheorem{remark}{Remark}

\newtheorem{assumption}{Assumption}

\def\Y{{\bf Y}}

\def\0{{\bf 0}}
\def\1{{\bf 1}}

\def\col{{\mathrm {col}}}
\def\st{{\mathrm {subject~to}}}

\def\eg{{\em e.g.}}
\def\ie{{\em i.e.}}

\def\diag{{\rm diag}}

\begin{document}
%
\title{
Implementation and Experimental Validation of Data-Driven Predictive Control for Dissipating Stop-and-Go Waves in Mixed Traffic
}

\author{Jiawei Wang, Yang Zheng, Jianghong Dong, Chaoyi Chen, Mengchi Cai, Keqiang Li and Qing Xu
	\thanks{The work of J. Wang, J. Dong, C. Chen, M. Cai, K. Li, and Q. Xu is supported by National Key R\&D Program of China with 2021YFB1600402, National Natural Science Foundation of China with 52072212, and Tsinghua-Toyota Joint Research Institute Cross-discipline Program. Corresponding author: Q.~Xu.}
	\thanks{J.~Wang, J. Dong, C. Chen, M. Cai, K. Li, and Q. Xu are with the School of Vehicle and Mobility, Tsinghua University, Beijing 100084, China. (\{wang-jw18,djh20,chency19,cmc18\}@mails.tsinghua.edu.cn, \{likq, qingxu\}@tsinghua.edu.cn).}%
	\thanks{Y. Zheng is with the Department of Electrical and Computer Engineering, University of California San Diego, CA 92093, USA. ({zhengy@eng.ucsd.edu}).}%
}

\maketitle

\begin{abstract}
	
In this paper, we present the first experimental results of data-driven predictive control for connected and autonomous vehicles (CAVs) in dissipating traffic waves. In particular, we consider a recent strategy of Data-EnablEd Predicted Leading Cruise Control (\method{DeeP-LCC}), 
which bypasses the need of identifying the driving behaviors of surrounding vehicles and directly relies on measurable traffic data to achieve safe and optimal CAV control in mixed traffic. We present the implementation details of \method{DeeP-LCC}, including data collection, equilibrium estimation, and control execution. Based on a miniature experiment platform, we reproduce the phenomenon of stop-and-go waves in two typical traffic scenarios: 1) open straight-road scenario under external disturbances and 2) closed ring-road scenario with no bottlenecks. Our experiments clearly demonstrate that \method{DeeP-LCC} enables one or a few CAVs to dissipate the traffic waves in both traffic scenarios. These experimental findings validate the great potential of \method{DeeP-LCC} in smoothing practical traffic flow in the presence of noisy data, uncertain low-level vehicle dynamics, and communication and computation delays. The code and videos of our experimental results are available at {\small \url{https://github.com/soc-ucsd/DeeP-LCC}}.
	
\end{abstract}


\section{Introduction}

\IEEEPARstart{T}{raffic} instabilities propagating upstream traffic flow can lead to periodical acceleration and deceleration of individual vehicles. This phenomenon is also known as phantom traffic jams when there are no apparent causes. The extreme stop-and-go traffic pattern further results in a significant societal loss of travel efficiency, fuel economy, and traffic safety~\cite{treiber2013traffic}. 
The seminal ring-road experiment in~\cite{sugiyama2008traffic} has shown that the traffic jams can be induced purely by the collective dynamics of human drivers without any external bottlenecks such as intersections, on/off ramps, and lane changes. Thanks to the rapid advances of vehicular communication and self-driving technologies, the emergence of connected and autonomous vehicles (CAVs) promises to revolutionize road transportation and significantly mitigate undesired traffic jams~\cite{guanetti2018control}. It has been shown that a full-CAV traffic system
 contributes to a dramatic increase in travel efficiency and safety guarantees~\cite{li2017dynamical,zheng2016stability,milanes2013cooperative}. 

The near future will meet an era of mixed traffic flow with the coexistence of both human-driven vehicles (HDVs) and CAVs. In such a human-in-the-loop mixed-autonomy system, dissipating traffic waves via CAVs is a promising strategy and has received increasing research interest. Despite the fact that HDVs cannot be directly controlled, a series of recent theoretical and experimental studies have revealed that CAVs promise significant traffic performance improvements, even in a low penetration rate, by incorporating the HDVs' behavior into controller design~\cite{stern2018dissipation,zheng2020smoothing,wang2020controllability,cui2017stabilizing,wu2018stabilizing,wang2021leading,orosz2016connected}. For example, the field tests in~\cite{stern2018dissipation} have empirically demonstrated the potential of one single CAV in dissipating stop-and-go waves in the ring-road setup of~\cite{sugiyama2008traffic}. CAVs' wave-dampening ability  in mixed traffic has also been validated via rigorous control-theoretic analysis, including controllability and stabilizability~\cite{zheng2020smoothing,wang2020controllability}, and head-to-tail string stability~\cite{wu2018stabilizing,orosz2016connected}. More recently, a notion of Leading Cruise Control (LCC)~\cite{wang2021leading,wang2020leadingCDC} has provided further insights into the role of CAVs in mixed traffic on common open straight-road scenarios. In the LCC framework, CAVs can not only \emph{adapt to} the downstream traffic flow (consisting of vehicles ahead), but also \emph{actively lead} the motion of upstream traffic flow (consisting of vehicles behind), contributing to system-wise improvements for the global traffic flow~\cite{wang2021leading}.

To realize the full potential of CAVs, it is important to design effective CAV control strategies in mixed traffic. 
One major challenge is how to incorporate the dynamics of HDVs, where human behaviors are known to be uncertain and stochastic. 
Most existing studies rely on HDVs' microscopic car-following models~\cite{wilson2011car}, and design control strategies for CAVs based on the corresponding dynamical model of mixed traffic systems~\cite{wang2020controllability,di2019cooperative,jin2017optimal,feng2021robust}. 
However, these model-based strategies may not be easily deployed in practice since it is non-trivial to accurately identify human driving behaviors for the car-following models. On the other hand, model-free or data-driven strategies~\cite{recht2019tour,furieri2020learning} can be used to stabilize mixed traffic without requiring explicit HDV's car-following models \textit{a priori}. With large-scale driving data of HDVs, some studies have proposed to train CAV control strategies for dampening traffic waves via reinforcement learning~\cite{wu2021flow} and adaptive dynamic programming~\cite{huang2020learning}. Yet, the requirement of extensive computational resources and the lack of safety guarantees in~\cite{wu2021flow,huang2020learning} remain practical limitations for real-world implementation. 

\begin{figure}[t]
	\centering
	\subfigure[Straight-road experiment]
	{
	\includegraphics[width=7.8cm]{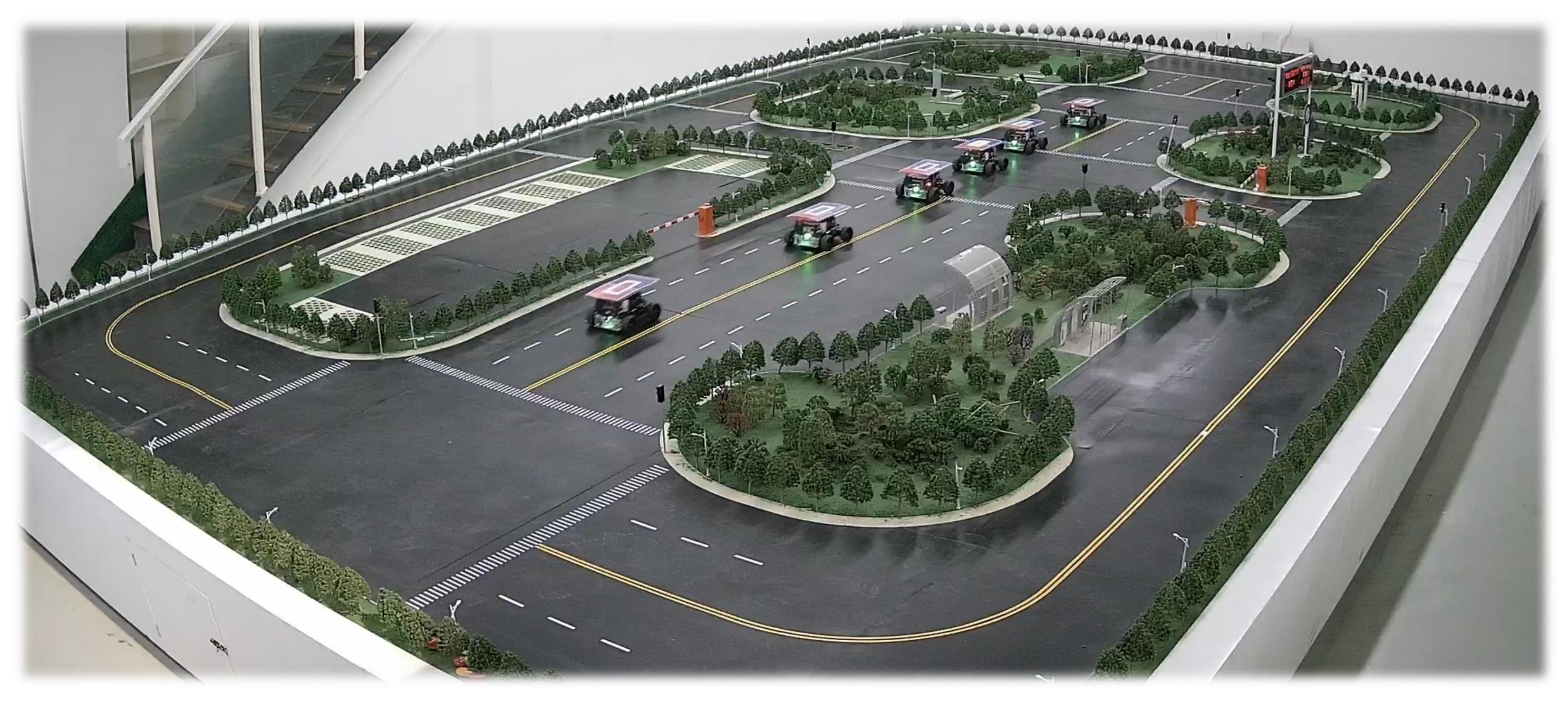}
	\label{Fig:Scenario_Straight}
	}
	\subfigure[Ring-road experiment]
	{
	\includegraphics[width=7.8cm]{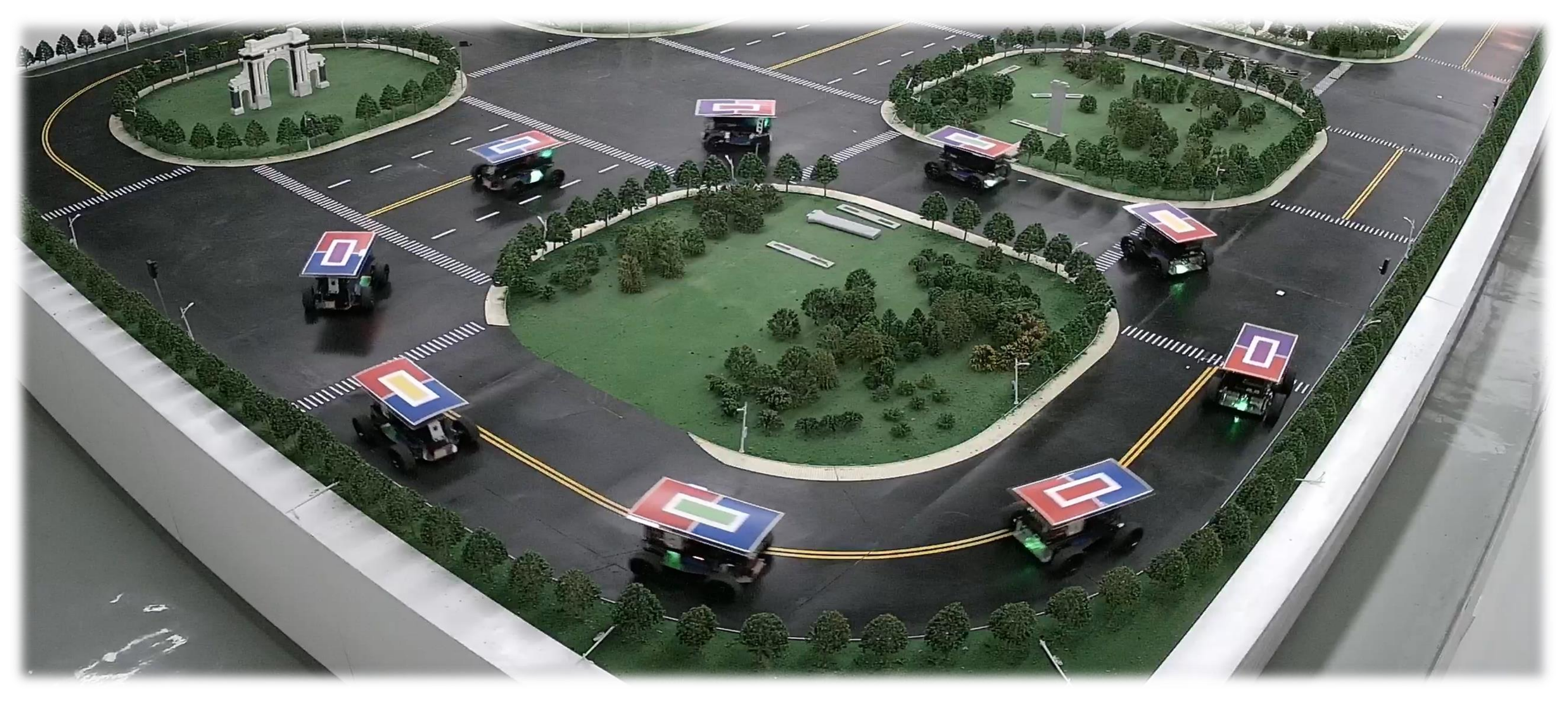}
	\label{Fig:Scenario_Ring}
	}
	\vspace{-2mm}
	\caption{Overview of the miniature experiments. Two main scenarios are under consideration: (a) Open straight-road scenario, where all the vehicles drive in a fleet with the head vehicle assigned a prescribed velocity trajectory. (b) Closed ring-road scenario, where each vehicle follows its predecessor. The code and videos of our experiments can be found in \url{https://github.com/soc-ucsd/DeeP-LCC}.}
	\label{Fig:Scenario}
	\vspace{-5mm}
\end{figure}

Data-driven predictive control has a great potential to resolve the aforementioned issues~\cite{hewing2020learning,markovsky2021behavioral}. 
Particularly, the recently proposed Data-EnablEd Predictive Leading Cruise Control (\method{DeeP-LCC}) strategy, which combines the Data-EnablEd Predictive Control (DeePC)~\cite{coulson2019data} with the LCC framework~\cite{wang2021leading}, promises to improve traffic performance with collision-free guarantees~\cite{wang2022data,wang2022deeplcc}. \method{DeeP-LCC} enables the CAVs to directly utilize measurable traffic data to obtain wave-dampening strategies, where input/output constraints are incorporated to avoid collision and actuation saturation. Rigorous controllability/observability analysis in~\cite{wang2022deeplcc} has provided theoretical insights for \method{DeeP-LCC}, and the
extensive simulations therein have confirmed the potential of \method{DeeP-LCC} in mitigating traffic waves and reducing fuel consumption while ensuring safety. To the best of our knowledge, however, all the existing \emph{data-driven techniques} for stabilizing traffic flow (\eg,~\cite{wu2021flow,huang2020learning,wang2022data,wang2022deeplcc,lan2021data}) have only been verified via virtual traffic simulations. It remains unclear about the performance of these data-driven techniques for CAV control using real-world noisy data in a practical setup. 

\subsection{Contributions}

In this paper, we provide the \textit{first} experimental results to validate the performance of data-driven predictive control in dissipating traffic waves. Precisely, a recent data-driven technique for CAV control, namely, \method{DeeP-LCC}, is under consideration. As shown in Fig.~\ref{Fig:Scenario}, we carry out real-world experiments on a miniature traffic platform deployed with robotic mini-vehicles. Compared to field tests with multiple real vehicles in~\cite{stern2018dissipation,brunner2022comparing,zheng2022empirical,gunter2020commercially}, miniature experiments have indeed received increasing attention for validating CAV technologies due to its greater flexibility, higher scalability, and simpler reproducibility, with lower labor and material costs; see, \eg,~\cite{hyldmar2019fleet,beaver2020demonstration,jang2019simulation}. Meanwhile, real-world factors such as vehicle dynamics and communication/computation delay can be naturally integrated in the miniature experiments, rather than being artificially generated based on pre-designed models in the simulation-related work~\cite{wu2021flow,huang2020learning,wang2022data,wang2022deeplcc,lan2021data}. 
In our experiments, we consider two specific traffic scenarios: open straight-road scenario and closed ring-road scenario, both of which are common in mixed traffic studies~\cite{sugiyama2008traffic,stern2018dissipation,wu2018stabilizing,orosz2016connected}. The robotic mini-vehicles are categorized into two types: HDVs and CAVs. A car-following model is utilized to capture the behavior of HDVs, and \method{DeeP-LCC} is implemented for CAVs using real driving data collected from the platform. In particular, the contributions of this paper are as follows. 
\begin{itemize}
    \item First, 
our experimental results are the first one that validate data-driven predictive control for dissipating traffic waves in a real-world environment. We provide a detailed implementation methodology of \method{DeeP-LCC}, including pre-collecting traffic data, estimating traffic equilibrium, and executing predictive control, which is capable of interacting with practical traffic flow and dealing with real-world trajectory data.
\item Second, going beyond the simulation setup in~\cite{wang2022deeplcc}, we consider the influence of CAV penetration rates and spatial locations of CAVs to demonstrate the applicability and benefits of \method{DeeP-LCC}. Compared with numerical simulations~\cite{wu2021flow,huang2020learning,wang2022data,lan2021data}, our experiments naturally integrate multiple practical factors of real-world operating features, including \textit{noisy measurement}, \textit{uncertain low-level vehicle dynamics} and \textit{real-world communication and computation delays}. Our experimental results reveal that \method{DeeP-LCC} overcomes the implementation challenges brought by these practical factors.
\item Third, our experiments reproduce the phenomenon of stop-and-go traffic waves not only in the straight-road scenario under external disturbances, but also in the ring-road scenario with no bottlenecks. Unlike existing empirical studies which focus on one single scenario (\eg,~\cite{stern2018dissipation,brunner2022comparing,zheng2022empirical,gunter2020commercially}), the capability of \method{DeeP-LCC} in dissipating traffic waves is validated in both scenarios, which reveals the generalization and flexibility property of data-driven predictive control. These experimental results validate the potential of data-driven techniques, particularly \method{DeeP-LCC}, for dissipating stop-and-go waves via CAVs. 
\end{itemize} 

The rest of this paper is organized as follows. Section~\ref{Sec:2} reviews the basics of \method{DeeP-LCC}. Section~\ref{Sec:3} introduces the miniature platform and experiment design, and the implementation of \method{DeeP-LCC} is detailed in Section~\ref{Sec:4}. Experimental results are presented in Section~\ref{Sec:5}.  Section~\ref{Sec:6} concludes this paper. 

\begin{figure*}[t]
	\centering
	\includegraphics[width=14cm]{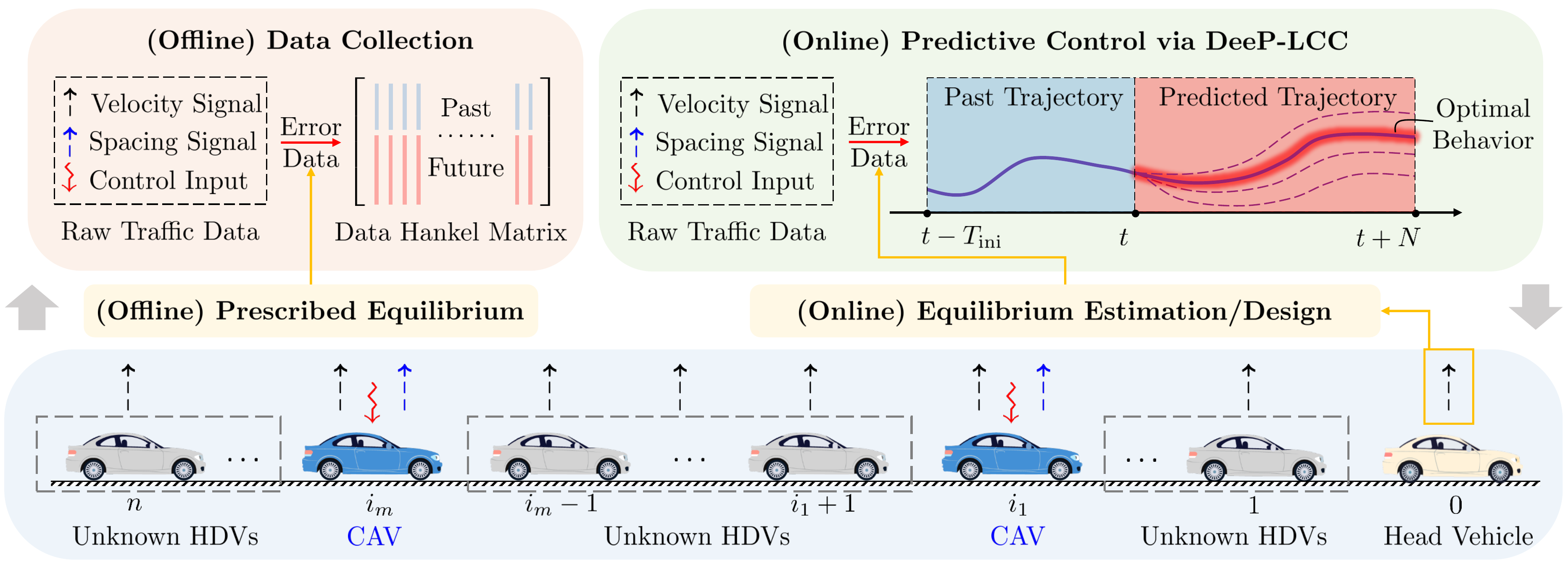}
	\vspace{-3mm}
\caption{Schematic for \method{DeeP-LCC}. The head vehicle is indexed as 0, behind which there exist $n$ vehicles consisting of $m$ CAVs and $n-m$ HDVs with unknown driving dynamics. \method{DeeP-LCC} is deployed in a cloud server, which is responsible for data collection, equilibrium estimation/design, and predictive control via \method{DeeP-LCC}. The collected data include velocity signals of all the vehicles (represented by black arrows), spacing signals of the CAVs (represented by blue arrows), and control input signals of the CAVs (represented by red arrows).}
	\label{Fig:SystemSchematic}
	\vspace{-4mm}
\end{figure*}

\section{Overview of \method{DeeP-LCC} for Mixed Traffic Flow}
\label{Sec:2}

In this section, we review the basics of the \method{DeeP-LCC} strategy~\cite{wang2022data,wang2022deeplcc} for CAV control in mixed traffic flow.

\subsection{Input/Output Data for Mixed Traffic Control}

We consider a general mixed traffic system, where there exist $n+1$ vehicles ($n \in \mathbb{N}$), including one head vehicle, indexed as $0$, and $n$ following vehicles, indexed from $1$ to $n$. Among the $n$ following vehicles, we have $m$ CAVs ($m \in \mathbb{N}$, $m \le n$) and $n-m$ HDVs. The behaviors of the CAVs can be controlled directly, while the HDVs' car-following dynamics are fixed but unknown. The schematic of the mixed traffic system is shown in Fig.~\ref{Fig:SystemSchematic}.  
Define $\Omega=\{1,2,\ldots,n\}$ as the set of all the vehicle indices, and $S=\{i_1,i_2,\ldots,i_m\}\subseteq \Omega$ ($i_1 < i_2 < \ldots < i_m$) as the set of the CAV indices. Note that the cardinality of $S$, \ie, $\vert S \vert = m$, represents the number of the CAVs, and thus $m/n$ corresponds to the CAV penetration rate, while the specific elements in $S$ represent the CAVs' spatial locations in traffic flow~\cite{li2022cooperative}. We denote $s_i(t)$, $v_i(t)$ and $a_i(t)$ as the inter-vehicle spacing, velocity and acceleration of vehicle $i$ at time $t$. 
 
The main objective for CAV control in mixed traffic is to stabilize traffic flow at a certain equilibrium, where each vehicle moves with an identical equilibrium velocity $v^*$ and a corresponding equilibrium spacing. The particular value of $v^*$ can be deduced from the steady-state behavior of the head vehicle~\cite{jin2018experimental}, while the equilibrium spacing for the HDVs is practically non-trivial to estimate and could be even time-varying. In contrast, the equilibrium spacing of CAVs can be designed, which is denoted as $s^*$. For notational simplicity, we consider a homogeneous setup for the equilibrium spacing of the CAVs, but all the results can be easily generalized to the heterogeneous scenario. The specific design of $s^*$ will be detailed in Section~\ref{Sec:4}. 

We next specify the input/output definition in mixed traffic. 
Similar to~\cite{zheng2020smoothing,jin2017optimal,huang2020learning}, the control input signal $u_i(t)$ ($i\in S$) of each CAV is assumed to be the acceleration $a_i(t)$, \ie, $a_i(t)=u_i(t)$. Lumping the control inputs of all the CAVs yields the aggregate control input of the entire mixed traffic: 
    \begin{equation} \label{Eq:ControlInput}
	u(t) = \begin{bmatrix}u_{i_{1}}(t), u_{i_{2}}(t), \ldots,         u_{i_{m}}(t)\end{bmatrix}^{\top} \in \mathbb{R}^{m}.
    \end{equation}
Meanwhile, an external input signal $\epsilon(t)\in \mathbb{R}$ is also introduced for the mixed traffic system, given by 
\begin{equation}
    \epsilon(t)=\tilde{v}_{0}(t)=v_0(t)-v^*,
\end{equation}
which is the velocity error of the head vehicle from the equilibrium velocity $v^*$. 
    
For the system output, we define the velocity error of each vehicle from the equilibrium as $\tilde{v}_i(t)=v_i(t)-v^*$ ($i \in \Omega$), and the spacing error of each CAV from the equilibrium as $\tilde{s}_i(t)=s_i(t)-s^*$ ($i \in S$). These measurable signals are lumped into the aggregate output of the mixed traffic system:  
    \begin{equation}\label{Eq:SystemOutput}
    y(t)=\begin{bmatrix}\tilde{v}_{1}(t),\tilde{v}_{2}(t),\ldots,\tilde{v}_{n}(t),\tilde{s}_{i_1}(t),\tilde{s}_{i_2}(t),\ldots,\tilde{s}_{i_m}(t)\end{bmatrix}^{\top},
    \end{equation}
where $y(t) \in \mathbb{R}^{n+m}    $.
Note that most existing studies, including~\cite{zheng2020smoothing,di2019cooperative,jin2017optimal,huang2020learning,gao2016data}, consider  state-feedback control by utilizing the velocity errors and spacing errors of all the vehicles in Fig.~\ref{Fig:SystemSchematic}. They assume that the state vector of mixed traffic, defined as 
\begin{equation} \label{Eq:SystemState}
    x(t)=\begin{bmatrix}\tilde{v}_{1}(t),\tilde{s}_{1}(t),\tilde{v}_{2}(t),\tilde{s}_{2}(t),\ldots,\tilde{v}_{n}(t),\tilde{s}_{n}(t)\end{bmatrix}^{\top},
\end{equation}
where $x(t)\in \mathbb{R}^{2n}$, is directly measurable.
This is impractical since the equilibrium spacing of the HDVs is unknown, and thus the spacing errors of the HDVs, \ie, $\tilde{s}_i(t)$ ($i \in \Omega \backslash S$), cannot be acquired in practice. In contrast, \method{DeeP-LCC} only employs measurable traffic data $y(t)$ in \eqref{Eq:SystemOutput} for control design.

\subsection{Non-Parametric Representation of Mixed Traffic Behavior}

Model-based strategies for mixed traffic control typically establish a parametric mixed traffic model for controller design. In particular, after specifying the system input, output and state in~\eqref{Eq:ControlInput}-\eqref{Eq:SystemState}, a discrete-time state-space model for the mixed traffic can be derived, which is in the form of~\cite{wang2022data,wang2022deeplcc}
\begin{equation} \label{Eq:DT_TrafficModel}
\begin{cases}
x(k+1) = A_\mathrm{d}x(k) + B_\mathrm{d}u(k) + H_\mathrm{d} \epsilon(k),\\
y(k) = C_\mathrm{d}x(k),
\end{cases}
\end{equation}
where $A_\mathrm{d},B_\mathrm{d},C_\mathrm{d},H_\mathrm{d}$ are system matrices of compatible dimensions. In~\cite{zheng2020smoothing,di2019cooperative,jin2017optimal}, typical car-following models, \eg, IDM~\cite{kesting2010enhanced} and OVM~\cite{bando1995dynamical}, are utilized for dynamical modeling of HDVs, and longitudinal vehicle dynamics are incorporated for CAVs. After linearization around the equilibrium, lumping the dynamics of all the HDVs and CAVs yields the mixed traffic system model~\eqref{Eq:DT_TrafficModel}; see~\cite[Section II]{wang2022deeplcc} for an exact form under the LCC framework. 

However, the car-following dynamics for individual HDVs are non-trivial to identify, and thus the system matrices  $A_\mathrm{d},B_\mathrm{d},C_\mathrm{d},H_\mathrm{d}$  in the parametric model~\eqref{Eq:DT_TrafficModel} are unknown in practice. Instead of using the parametric model~\eqref{Eq:DT_TrafficModel}, \method{DeeP-LCC} relies on Willems' fundamental lemma~\cite{willems2005note} to establish a data-centric non-parametric representation using traffic data in~\eqref{Eq:ControlInput}-\eqref{Eq:SystemOutput}. 
For notations, given a collection of vectors $a_1,a_2,\ldots,a_m$, we denote $\col(a_1,a_2,\ldots,a_m)=\begin{bmatrix}
a_1^\top,a_2^\top,\ldots,a_m^\top
\end{bmatrix}^\top$.

\begin{definition}
	The signal  $\omega = \col \left(\omega(1),\omega(2),\ldots,\omega(T)\right)$ of length $T$ $(T\in \mathbb{N})$ is persistently exciting of order $l$ $(l\in \mathbb{N}, l \leq T)$, if the following Hankel matrix is of full row rank
	\begin{equation}
		\mathcal{H}_{l}(\omega):=\begin{bmatrix}
			\omega(1) &\omega(2) & \cdots & \omega(T-l+1) \\
			\omega(2) &\omega(3) & \cdots & \omega(T-l+2) \\
			\vdots & \vdots & \ddots & \vdots \\
			\omega(l) &\omega(l+1) & \cdots & \omega(T)
		\end{bmatrix}.
	\end{equation}
\end{definition}

\method{DeeP-LCC} begins by collecting a sequence of input/output data of length $T$ from the mixed traffic system~\eqref{Eq:DT_TrafficModel} with sampling time interval $\Delta t$, given as
\begin{equation} \label{Eq:CollectedData}
\begin{aligned}
u^\mathrm{d}&=\col \left(u^\mathrm{d}(1),u^\mathrm{d}(2),\ldots,u^\mathrm{d}(T)\right)\in \mathbb{R}^{mT},\\
\epsilon ^\mathrm{d}&=\col \left(\epsilon^\mathrm{d}(1),\epsilon^\mathrm{d}(2),\ldots,\epsilon^\mathrm{d}(T)\right) \in \mathbb{R}^{T},\\
y ^\mathrm{d}&=\col \left(y^\mathrm{d}(1),y^\mathrm{d}(2),\ldots,y^\mathrm{d}(T)\right) \in \mathbb{R}^{(n+m)T}.
\end{aligned}
\end{equation}
Then, these pre-collected data are partitioned into two parts: 1) \emph{past data} with a time horizon of $T_{\mathrm{ini}}$ ($T_{\mathrm{ini}} \in \mathbb{N}$) and 2) \emph{future data} with a time horizon of $N$ ($N \in \mathbb{N}$). Precisely, the data partition is as follows
\begin{equation}
\label{Eq:DataHankel}
\begin{gathered}
\begin{bmatrix}
U_{\mathrm{p}} \\
U_{\mathrm{f}}
\end{bmatrix}:=\mathcal{H}_{T_{\mathrm{ini}}+N}(u^{\mathrm{d}}), \quad \begin{bmatrix}
E_{\mathrm{p}} \\
E_{\mathrm{f}}
\end{bmatrix}:=\mathcal{H}_{T_{\mathrm{ini}}+N}(\epsilon^{\mathrm{d}}), \\
\begin{bmatrix}
Y_{\mathrm{p}} \\
Y_{\mathrm{f}}
\end{bmatrix}:=\mathcal{H}_{T_{\mathrm{ini}}+N}(y^{\mathrm{d}}),
\end{gathered}
\end{equation}
where $U_{\mathrm{p}}$ and $U_{\mathrm{f}}$ consist of the first $T_{\mathrm{ini}}$ block rows and the last $N$ block rows of $\mathcal{H}_{T_{\mathrm{ini}}+N}(u^{\mathrm{d}})$, respectively (similarly for $E_{\mathrm{p}}, E_{\mathrm{f}}$ and $Y_{\mathrm{p}}, Y_{\mathrm{f}}$). For the subsequent non-parametric behavior representation, we introduce two assumptions. 

\begin{assumption} \label{Assumption:PersistentExcitation}
The pre-collected combined input data, defined as $\hat{u}^\mathrm{d} = \col \left(u^\mathrm{d}(1),\epsilon^\mathrm{d}(1),\ldots,u^\mathrm{d}(T),\epsilon^\mathrm{d}(T) \right)$, is persistently exciting of order $T_\mathrm{ini}+N+2n$.
\end{assumption}

\begin{assumption} \label{Assumption:Controllable}
The mixed traffic system~\eqref{Eq:DT_TrafficModel} is controllable and observable by regarding $u(t)$ and $\epsilon(t)$ as a combined input $\hat{u}(t)$, \ie, $\hat{u}(t) = \col \left(\epsilon (t),u (t) \right)$.
\end{assumption}

The persistent excitation requirement in Assumption~\ref{Assumption:PersistentExcitation} can be easily satisfied in experiments given the nature of uncertain human behavior in ${\epsilon}^\mathrm{d}$ and an i.i.d. random setup of ${u}^\mathrm{d}$ in data collection. The controllability and observability of the mixed traffic system~\eqref{Eq:DT_TrafficModel} in Assumption~\ref{Assumption:Controllable} have been proved under a mild condition  in~\cite[Theorem 1]{wang2022deeplcc}. 
Motivated by Willems' fundamental lemma~\cite{willems2005note} and standard DeePC~\cite{coulson2019data}, we have the following result for \method{DeeP-LCC}~\cite{wang2022deeplcc}.

\begin{proposition}
\label{Proposition:DeePCMixedTraffic}
At the time step $t$, we define
\begin{equation} \label{Eq:PastDataDefinition}
    \begin{aligned}
    u_{\mathrm{ini}}&=\col\left(u(t-T_{\mathrm{ini}}),u(t-T_{\mathrm{ini}}+1),\ldots,u(t-1)\right),\\
    u&= \col\left(u(t),u(t+1),\ldots,u(t+N-1)\right),
\end{aligned}
\end{equation} 
as the past control input sequence of length $T_{\mathrm{ini}}$ and the future control input sequence of length $N$, respectively (similarly for $\epsilon_\mathrm{ini},\epsilon$ and $y_\mathrm{ini},y$). Under Assumptions~\ref{Assumption:PersistentExcitation} and~\ref{Assumption:Controllable}, any length-$(T_{\mathrm{ini}}+N)$ trajectory of the mixed traffic system~\eqref{Eq:DT_TrafficModel}, denoted as $\col (u_\mathrm{ini},\epsilon_\mathrm{ini},y_\mathrm{ini},u,\epsilon,y)$, can be constructed as
\begin{equation}
\label{Eq:AdaptedDeePCAchievability}
\begin{bmatrix}
U_\mathrm{p} \\ E_\mathrm{p}\\Y_\mathrm{p} \\ U_\mathrm{f} \\ E_\mathrm{f}\\ Y_\mathrm{f}
\end{bmatrix}g=
\begin{bmatrix}
u_\mathrm{ini} \\ \epsilon_\mathrm{ini}\\ y_\mathrm{ini} \\ u \\\epsilon \\ y
\end{bmatrix},
\end{equation}
where $g\in \mathbb{R}^{T-T_\mathrm{ini}-N+1}$. Furthermore, if $T_{\mathrm{ini}} \geq 2n$, $y$ is uniquely determined from~\eqref{Eq:AdaptedDeePCAchievability}, $\forall (u_\mathrm{ini} ,\epsilon_\mathrm{ini}, y_\mathrm{ini},u,\epsilon)$. 
\end{proposition}

Willems' fundamental lemma~\cite{willems2005note} reveals that for a controllable linear time-invariant (LTI) system, the subspace consisting of all valid trajectories is identical to the range space of data Hankel matrices of the same order generated by sufficiently rich inputs. \method{DeeP-LCC} applies this result to mixed traffic control and derives the representation~\eqref{Eq:AdaptedDeePCAchievability}, which allows one to use past input/output data to predict the future input/output behavior of the mixed traffic system without explicitly identifying a parametric mixed traffic model~\eqref{Eq:DT_TrafficModel}. 



\subsection{Formulation of \method{DeeP-LCC}}

The non-parametric representation~\eqref{Eq:AdaptedDeePCAchievability} can be utilized for predictive control in mixed traffic. For the linearized mixed traffic system~\eqref{Eq:DT_TrafficModel} with noise-free data~\eqref{Eq:CollectedData}, we formulate the following optimization problem 
\begin{equation} \label{Eq:AdaptedDeePC}
\begin{aligned}
\min_{g,u,y} \quad &J(y,u) \\
\st \quad &\eqref{Eq:AdaptedDeePCAchievability},\\
&\epsilon = \hat{\epsilon}, \\
 & u\in \mathcal{U},y\in \mathcal{Y},
\end{aligned}
\end{equation}
where $J(y,u)$ denotes a cost function, and $u\in \mathcal{U},y\in \mathcal{Y}$ represents  input/output constraints. Note that the future external input sequence $\epsilon$ cannot be designed; instead, we use $\hat{\epsilon}$ to represent its future estimation in \eqref{Eq:AdaptedDeePC}.  At the time step $t$, given past trajectory $(u_\mathrm{ini},\epsilon_\mathrm{ini},y_\mathrm{ini})$, solving~\eqref{Eq:AdaptedDeePC} offers the optimal future trajectory $(u,y)$.

Problem \eqref{Eq:AdaptedDeePC} is valid for the linear time-invariant mixed traffic system \eqref{Eq:DT_TrafficModel} with noise-free data. In practical traffic flow, the HDVs' car-following behavior is nonlinear and non-deterministic, and the collected data from sensors are also noise-corrupted. In this case, the non-parametric behavior representation~\eqref{Eq:AdaptedDeePCAchievability} is not consistent, and thus the optimization problem~\eqref{Eq:AdaptedDeePC} might have no feasible solutions. To deal with nonlinear and non-deterministic mixed traffic flow, a regularized version of~\eqref{Eq:AdaptedDeePC} is proposed to compute the optimal control input in \cite{wang2022deeplcc} , given as follows
  \begin{equation} \label{Eq:AdaptedDeePCforNonlinearSystem}
 \begin{aligned}
 \min_{g,u,y,\sigma_y} \quad &J(y,u)+\lambda_g \left\|g\right\|_2^2+\lambda_y \left\|\sigma_y\right\|_2^2\\
 \st \quad & \begin{bmatrix}
 U_\mathrm{p} \\ E_\mathrm{p}\\Y_\mathrm{p} \\ U_\mathrm{f} \\ E_\mathrm{f}\\ Y_\mathrm{f}
 \end{bmatrix}g=
 \begin{bmatrix}
 u_\mathrm{ini} \\ \epsilon_\mathrm{ini}\\ y_\mathrm{ini} \\ u \\\epsilon \\ y
 \end{bmatrix}+\begin{bmatrix}
 0\\0\\ \sigma_y \\0 \\0 \\0
 \end{bmatrix},\\ &\epsilon = \hat{\epsilon},\\
 & u\in \mathcal{U},y\in \mathcal{Y},
 \end{aligned}
 \end{equation}
 which is the main optimization problem in  \method{DeeP-LCC} at each time step $t$. In~\eqref{Eq:AdaptedDeePCforNonlinearSystem},
a slack variable $\sigma_y \in \mathbb{R}^{(n+m)T_\mathrm{ini}}$ is introduced, and it is penalized with a weighted two-norm value in the regulated cost function. A sufficiently large value of the weight coefficient $\lambda_y>0$ allows for $\sigma_y \neq 0$ only if the equality constraint is infeasible. In addition, a two-norm penalty on $g$ with a weight coefficient $\lambda_g>0$ is also incorporated to avoid over-fitting of noisy data. This regulation has been shown to be consistent with distributional two-norm robustness~\cite{huang2019decentralized,coulson2019regularized}. More details on \method{DeeP-LCC} can be found in \cite{wang2022data,wang2022deeplcc}.

\section{Experimental Design for \method{DeeP-LCC} Validation}
\label{Sec:3}

In this section, we introduce our experimental platform and setup for the validation of \method{DeeP-LCC} in mixed traffic.

\begin{figure}[t]
	\vspace{1mm}
	\centering
	\includegraphics[width=8cm]{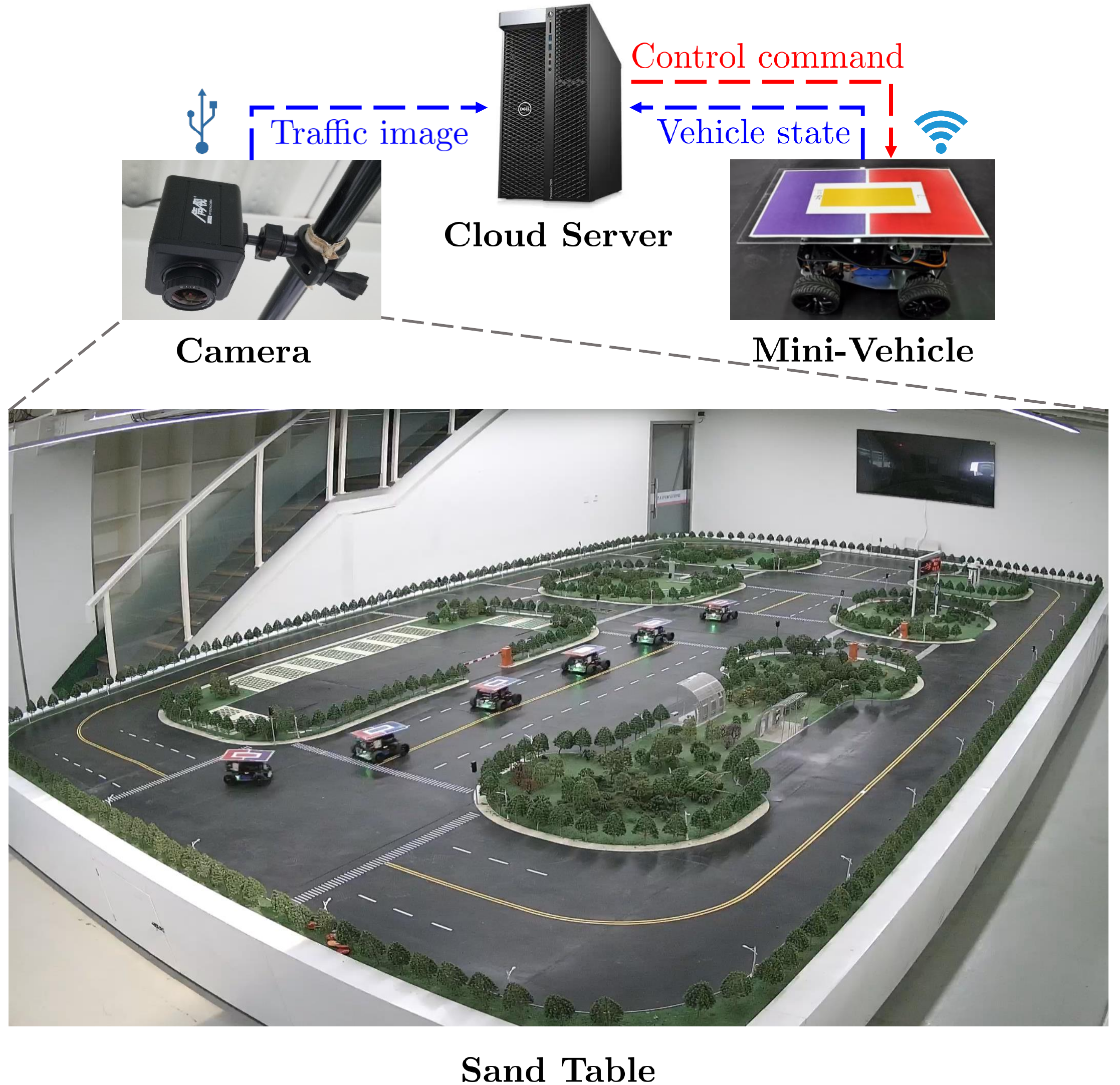}
	\vspace{-2mm}
	\caption{Schematic for the architecture of the experimental platform. The cameras capture a top view image of the road network on the sand table and transmit the image to the cloud server via USB cables. The mini-vehicles on the sand table transmit its state information to the cloud server via WiFi. The cloud server then integrates the received information and calculates the control command, which is then transmitted to each mini-vehicle.}
	\label{Fig:Platform_archecticture}
	\vspace{-2mm}
\end{figure}

\subsection{Experiment Platform}
\label{Sec:Platform}

\begin{figure}[t]
	\centering
	\includegraphics[width=7cm]{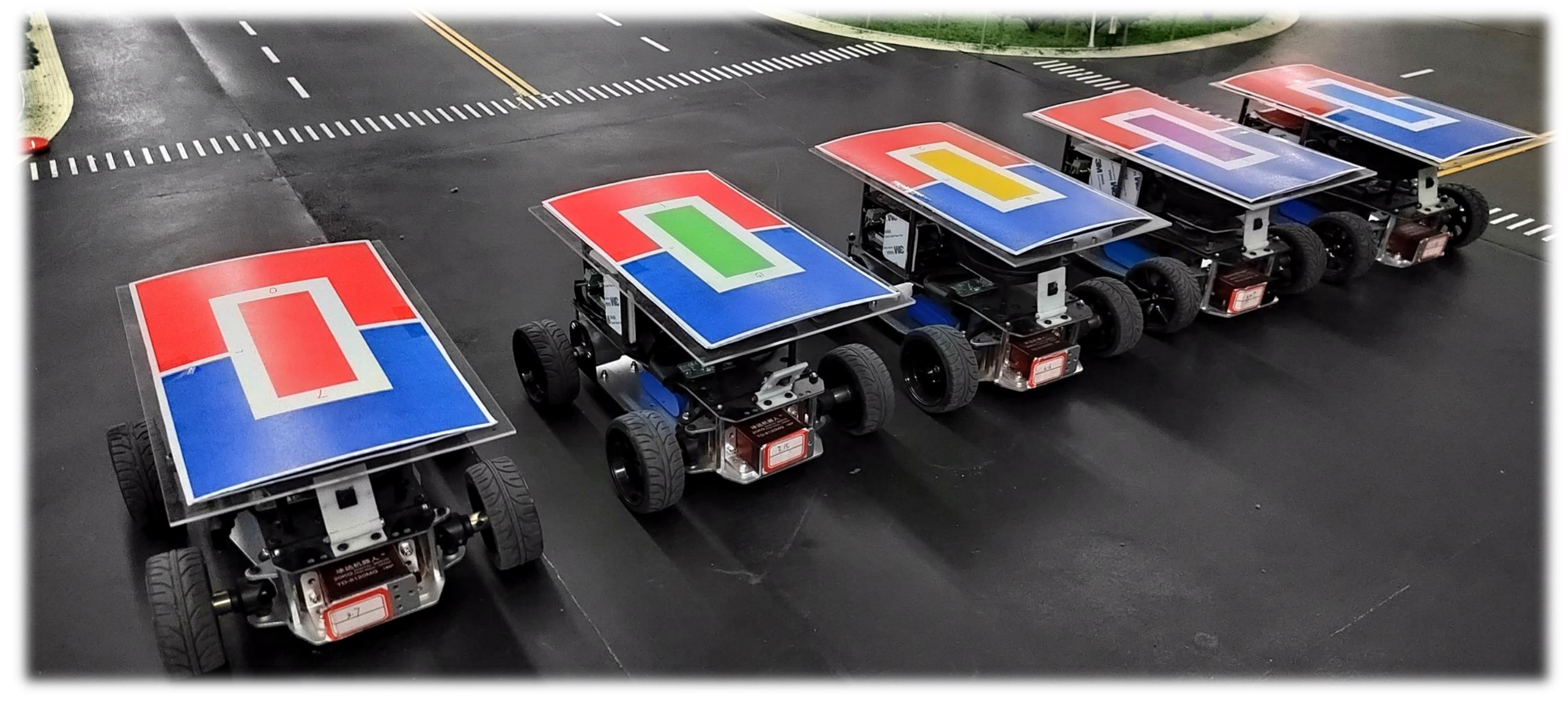}
	\vspace{-2mm}
	\caption{Experimental robotic mini-vehicles. The mini-vehicles receive velocity and steering angles commands from the cloud server, and have on-board actuators (motors and steering gears) to execute these commands. A piece of colored cardboard is sticked to the top of each vehicle for vehicle localization via cameras installed on the ceiling. }
	\label{Fig:Platform_vehicle}
	\vspace{-2mm}
\end{figure}

We test \method{DeeP-LCC} in a specially designed platform at Tsinghua University~\cite{yang2022multi}. The platform architecture is shown in Fig.~\ref{Fig:Platform_archecticture}. A sand table of miniature road network is constructed, whose size is $9\,\mathrm{m}\times 5\,\mathrm{m}$ and the lanes are of width $240\,\mathrm{mm}$. Commercially available robotic mini-vehicles (on-board CPU: Raspberry Pi 4 Model B) are deployed on the sand table and they are of size $200\,\mathrm{mm}\times 200\,\mathrm{mm} \times 130\,\mathrm{mm}$; see Fig.~\ref{Fig:Platform_vehicle} for illustration. A cloud server (CPU: Intel Core i7-10700K, GPU: NVIDIA GeForce RTX 2080 SUPER) collects platform data and computes the control commands for each mini-vehicle. Four cameras (focal length: $2.4\, \mathrm{mm}$, operating frame rate: $30 \, \mathrm{FPS}$, image resolution:  $1920 \times 1080$) are installed on the ceiling to take a real-time top view picture of the sand table for vehicle localization in the road network. We present the details of state measurement and control computation below. 

\subsubsection{State measurement} 

\begin{table}[t!]
\footnotesize
	\begin{center}
		\caption{Localization and WiFi communication delays}
		\vspace{-2mm}
		\label{Tb:Delay}
		\setlength{\tabcolsep}{2mm}{
		\begin{tabular}{ccc}
		\toprule
			& Mean Value & Standard Deviation \\\hline
			Localization Delay & $49.55\,\mathrm{ms}$ & $1.39\,\mathrm{ms}$  \\
			WiFi Communication Delay & $1.71\,\mathrm{ms}$ & $0.66\,\mathrm{ms}$  \\
			\bottomrule
		\end{tabular}}
	\end{center}
	\vspace{-4mm}
\end{table}

\begin{table}[t!]
\footnotesize
	\begin{center}
		\caption{Static Measurement Noise} 
		\vspace{-2mm}
		\label{Tb:Noise_Error}
		\begin{threeparttable}
		\setlength{\tabcolsep}{4mm}{
		\begin{tabular}{ccc}
		\toprule
			& Mean Value & Standard Deviation \\\hline
			Position.X & $2.28\,\mathrm{mm}$ & $1.77\,\mathrm{mm}$  \\
			Position.Y & $6.13\,\mathrm{mm}$ & $1.16\,\mathrm{mm}$  \\
			Heading Angle & $0.032\,\mathrm{rad}$ & $0.0159\,\mathrm{rad}$  \\
			Velocity & $-0.0010\,\mathrm{m/s}$ & $0.0054\,\mathrm{m/s}$  \\
			\bottomrule
		\end{tabular}}
		\begin{tablenotes}
		\footnotesize
		\item[1] Position.X and Position.Y represent the horizontal axis and the vertical axis in the top view in Fig.~\ref{Fig:Platform_archecticture}, respectively.
		\item[2] The mean value of measurement noise can also be regarded as an average measurement error.
		\end{tablenotes}
		\end{threeparttable}
	\end{center}
	\vspace{-4mm}
\end{table}

For each mini-vehicle, the measurable signals in the platform include velocity, position and heading angle. Specifically, the velocity signal is measured by the on-board sensors of each vehicle, and then transmitted to the cloud server via WiFi 6 (AX3 Pro WLAN Router). For localization, the real-time top view picture of the sand table taken by the cameras is transmitted to the cloud server via USB 3.1 cables. The cloud server then processes this picture to measure the position and heading angle of each vehicle in the road network by detecting the color cardboard installed on the vehicle top via OpenCV (see Fig.~\ref{Fig:Platform_vehicle} for the hardware layout of the mini-vehicles). It is inevitable that the whole system has  time delays of localization and WiFi communications. We have conducted $7500$ random tests to get statistics for these delays, which  are listed in Table~\ref{Tb:Delay}. The measurement noise of velocity, position and heading angle are also tested for a time period of $500\,\mathrm{s}$, and the results are listed in Table~\ref{Tb:Noise_Error}.

\subsubsection{Control computation} 

The executable control commands for the mini-vehicles include command velocity and command steering angle, which are computed and issued by the cloud server. The cloud server has a prescribed trajectory for each vehicle to follow, and the longitudinal control and lateral control are decoupled to derive the command velocity and steering angle, respectively. For the longitudinal control, either an HDVs' car-following model or \method{DeeP-LCC} is adopted, depending on the type of the vehicle (HDV or CAV). For the lateral control, a typical preview trajectory tracking controller is used~\cite{amer2017modelling}. These control commands are then transmitted to each vehicle via WiFi and executed by the on-board actuators (motors and steering gears).~As for the software, MATLAB R2021a is utilized for control computation in the cloud server, and ROS Melodic is employed for interaction between the server and the vehicles.  

For the longitudinal control of HDVs, we utilize a car-following model to capture HDVs' driving behavior, as motivated by existing experimental work on mixed traffic~\cite{beaver2020demonstration,jang2019simulation}. We consider the OVM model in our implementation~\cite{jin2017optimal,bando1995dynamical}, which is in the form of
\begin{equation} \label{Eq:OVMmodel}
a_i(t)=\alpha\left(v_{\mathrm{des}}\left(s_i(t)\right)-v_i(t)\right)+\beta\dot{s}_i(t),\quad i \in \Omega \backslash S,
\end{equation}
where $\dot{s}_i(t)=v_{i-1}(t)-v_i(t)$ denotes the relative velocity of vehicle $i$ at time $t$, and $\alpha, \beta > 0$ denote the sensitivity coefficients. We use $v_{\mathrm{des}}(s)$ to represent the spacing-dependent desired velocity for HDVs, given by a continuous piece-wise function
\begin{equation} \label{Eq:OVMDesiredVelocity}
v_{\mathrm{des}}(s)=\begin{cases}
0, &s\le s_{\mathrm{st}};\\
\displaystyle \frac{v_{\max }}{2}\left(1-\cos (\pi\frac{s-s_{\mathrm{st}}}{s_{\mathrm{go}}-s_{\mathrm{st}}})\right), &s_{\mathrm{st}}<s<s_{\mathrm{go}};\\
v_{\max}, &s\ge s_{\mathrm{go}},
\end{cases}
\end{equation}
where $v_{\max }$ denotes the maximum velocity, and $s_\mathrm{st},s_\mathrm{go}$ denote the stand-still spacing and free-driving spacing, respectively.

We note that the vehicle acceleration is regarded as the control signal in both the OVM model~\eqref{Eq:OVMmodel} and \method{DeeP-LCC} \eqref{Eq:AdaptedDeePCforNonlinearSystem}, but the executable longitudinal control command for the mini-vehicles in the platform is the velocity signal. To address this gap, the command acceleration $a_i(t)$ of each vehicle is translated into the command velocity signal $v_{i,\mathrm{cmd}}(t)$ in the cloud server by
\begin{equation} \label{Eq:CommandVelocity}
    v_{i,\mathrm{cmd}}(t) = v_{i}(t-1) + a_i(t) \Delta t _ {\mathrm{real}},
\end{equation}
where $v_{i}(t-1)$ denotes the measured velocity of vehicle $i$ at the last time step $t-1$, and $\Delta t _ {\mathrm{real}}$ denotes the real time interval between time $t$ and time $t-1$, which keeps updated during the real-time experiments.

\begin{figure}[t!]
	\vspace{2mm}
	\centering
	\includegraphics[width=7.5cm]{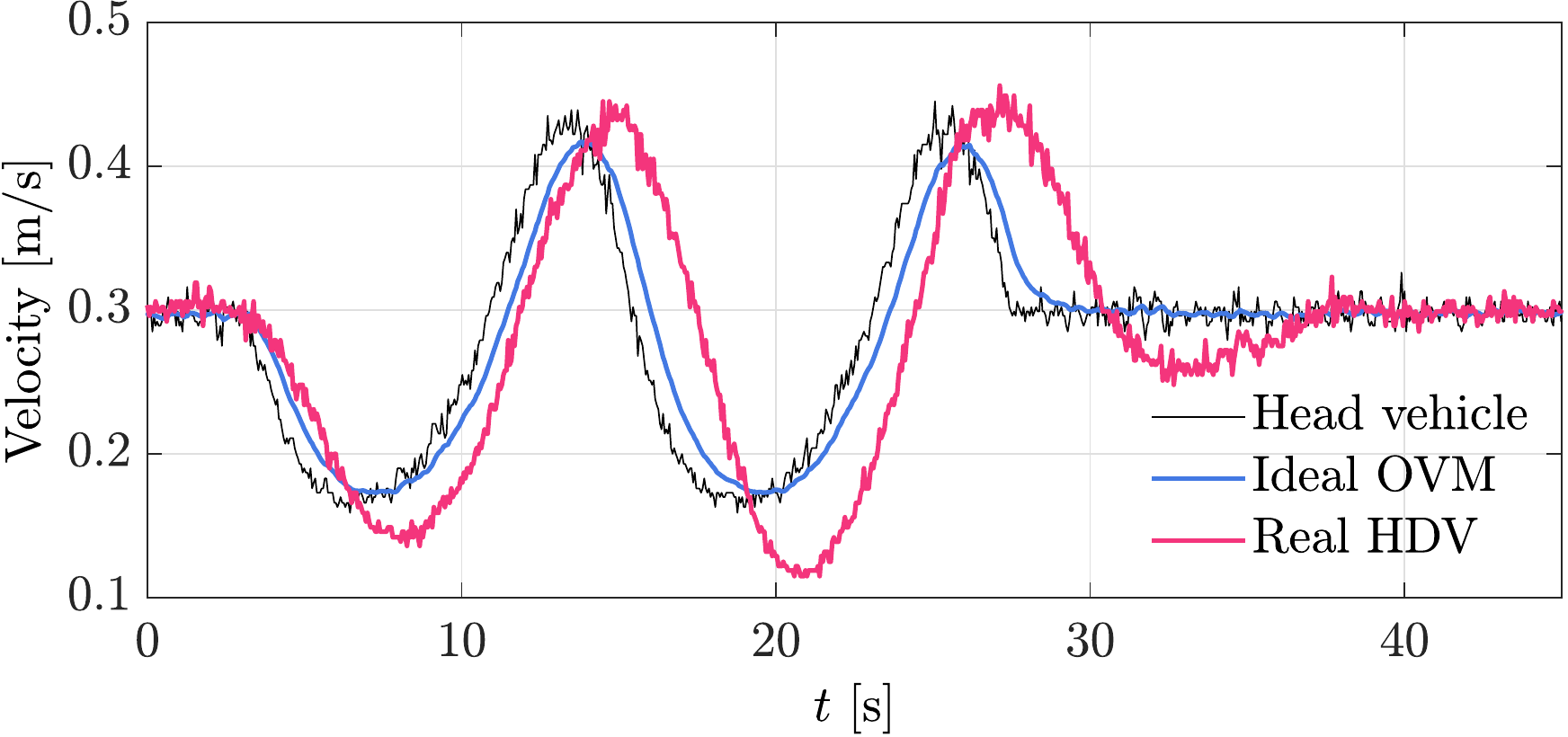}
	\vspace{-2mm}
	\caption{Real driving performance of HDVs under the OVM model~\eqref{Eq:OVMmodel} in the experiment platform. The OVM parameters are shown in Table~\ref{Tb:OVM}. The profile of ``Ideal OVM" represents the velocity trajectory directly from OVM with no low-level dynamics, while the profile of ``Real HDV" represents the real-world trajectory of the mini-vehicles in the experiments.}
	\label{Fig:OVM_trajectory}
	\vspace{-2mm}
\end{figure}

\begin{table}[t!]
\footnotesize
	\centering
		\caption{Parameter Setup for OVM}
		\vspace{-2mm}
		\label{Tb:OVM}
		\setlength{\tabcolsep}{2mm}{
		\begin{tabular}{ccc}
		\toprule
			Symbol & Meaning & Value \\\hline
			$s_\mathrm{st}$ & Spacing lower bound  & $0.5$  \\
			$s_\mathrm{go}$ & Spacing upper bound  & $1.1$\\
			$\alpha$ & Sensitivity coefficient  & $1.2$ (straight road) or $2.4$ (ring road) \\
			$\beta $ & Sensitivity coefficient  & $1.8$ (straight road) or $3.6$ (ring road)  \\
			$v_\mathrm{max}$ & Maximum velocity & $0.6$ \\
			\bottomrule
		\end{tabular}}
\end{table}

\begin{remark}[Uncertain lower-level dynamics]
Due to the updating mechanism~\eqref{Eq:CommandVelocity} of control commands and the original low-level dynamics of the mini-vehicles, the real driving behavior of the mini-vehicles shows some inconsistency with the ideal car-following model. Consider a straight-road scenario where the head vehicle has a sinusoid perturbation and one HDV with OVM following behind. The comparison results of the velocity trajectories between the ideal OVM model~\eqref{Eq:OVMmodel} (see Table~\ref{Tb:OVM} for parameter values) and the real-world driving behavior observed in the platform are shown in Fig.~\ref{Fig:OVM_trajectory}. This influence of uncertain low-level dynamics from the control command in the cloud server to the real driving behavior of the vehicles is naturally integrated and tested in our experiment platform. 
\end{remark}

\subsection{Experiment Design}

\label{Sec:ExperimentDesign}

\begin{figure*}[t]
	\centering
	\subfigure[straight-road experiment]
	{
	\includegraphics[width=8cm]{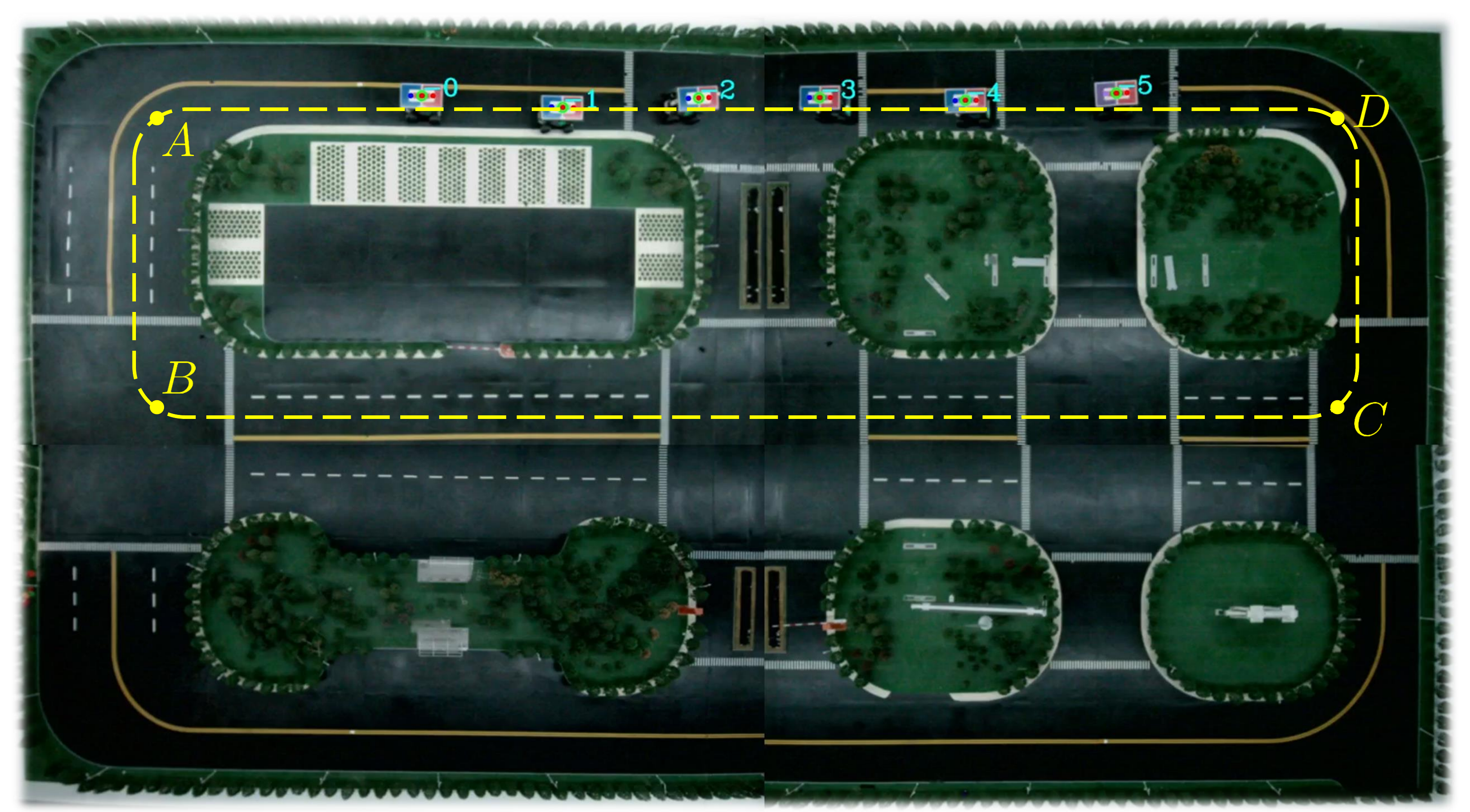}
	\label{Fig:Overview_Straight}
	}
	\subfigure[ring-road experiment]
	{
	\includegraphics[width=8cm]{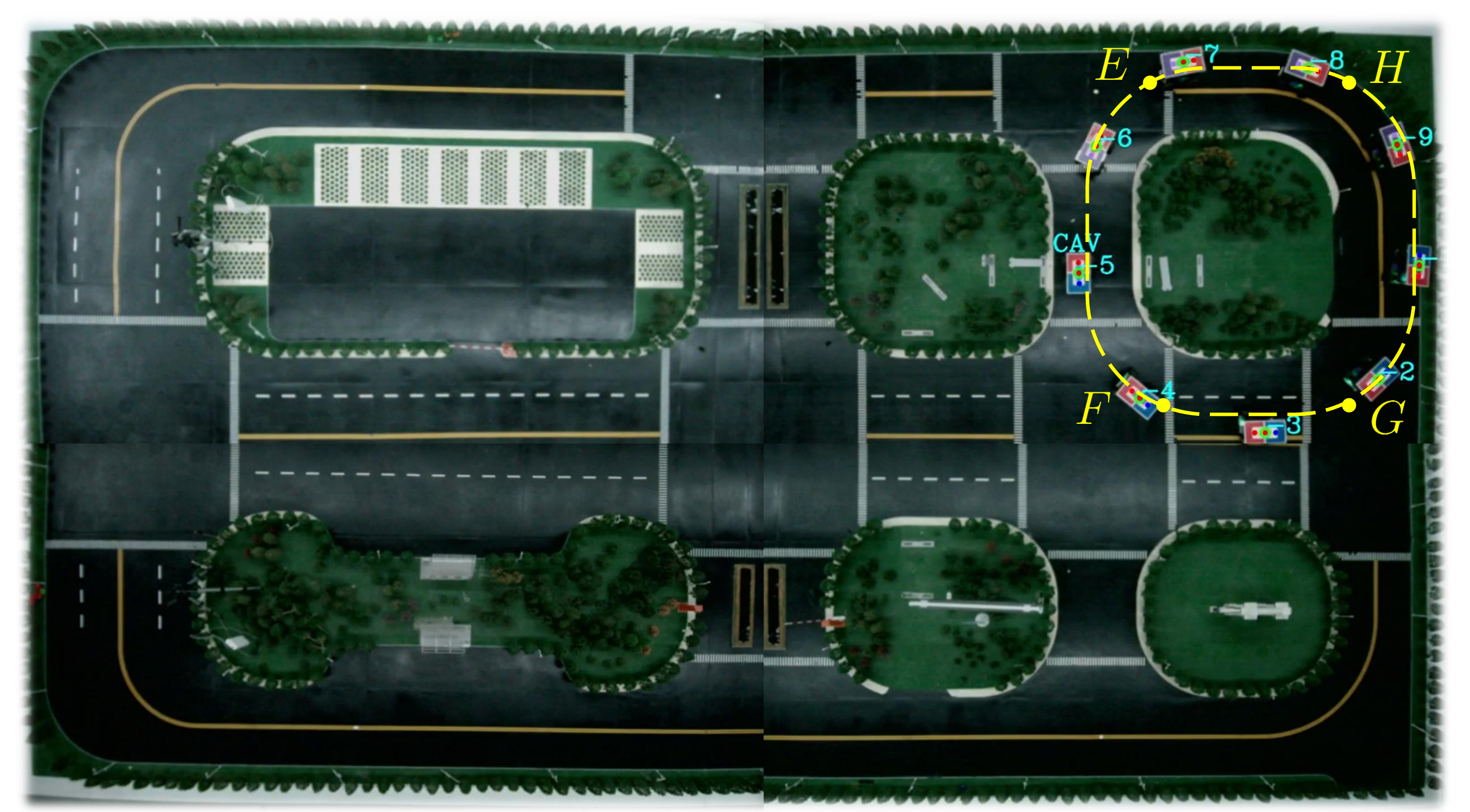}
	\label{Fig:Overview_Ring}
	}
	\vspace{-2mm}
	\caption{ Top view of the two experiments. The yellow dashed line represents the prescribed trajectory of each vehicle in the experiments. (a) In the experiments for the straight-road scenario, the vehicles drive along the direction $A \rightarrow B\rightarrow C\rightarrow D\rightarrow A$. (b) In the experiments for the ring-road scenario, the vehicles drive along the direction $E \rightarrow F\rightarrow G\rightarrow H\rightarrow E$.}
	\label{Fig:Overview}
	\vspace{-3mm}
\end{figure*}

In our experiments, we have considered two different traffic scenarios: straight-road experiment and ring-road experiment.

\subsubsection{Straight-road experiment} 

We first consider an open single-lane track of lap $17.5\,\mathrm{m}$ in the platform; see $A \rightarrow B\rightarrow C\rightarrow D\rightarrow A$ in Fig.~\ref{Fig:Overview_Straight}. A fleet of $6$ mini-vehicles running along this track is deployed in this experiment. Although there exist four bends where the vehicles need to turn (see the location at $A,B,C,D$), the longitudinal control strategies remain unchanged for each vehicle. Thus, this experiment is close to the scenario where a fleet of vehicles are driving along an infinite-length straight-road. 

Particularly, the velocity of the head vehicle (vehicle $0$ in Fig.~\ref{Fig:Overview_Straight}) is pre-designed. Similar to the simulation studies~\cite{feng2021robust,Vaio2019Cooperative} and experimental work~\cite{jin2018experimental,naus2010string}, the head vehicle suffers from a sinusoid perturbation when it drives along $B \rightarrow C$, while in other segments it maintains a constant velocity $v_\mathrm{c} = 0.3 \, \mathrm{m/s}$. This setup simulates the scenario where there exists an external disturbance on the road $B \rightarrow C$ (\emph{e.g.}, cut in, ramp in/out, and intersection), which causes the head vehicle to be caught in traffic waves. The designed velocity trajectory of the head vehicle is shown in Table~\ref{Tb:HeadVehicleTrajectory}.

\begin{table}[t]
\footnotesize
	\begin{center}
		\caption{Velocity of the Head Vehicle in straight-road Experiments}
		\vspace{-2mm}
		\label{Tb:HeadVehicleTrajectory}
		\begin{threeparttable}
		\setlength{\tabcolsep}{2mm}{
		\begin{tabular}{cc}
		\toprule
			Segment & Velocity Trajectory \\\hline
			$C \rightarrow D \rightarrow A \rightarrow B$ & $v_0 =  v_\mathrm{c}$ \\
			$B \rightarrow C$ & $v_0 = v_c - 0.13 \sin{(4 \pi d_\mathrm{B-0}/d_\mathrm{B-C} )}$\\
			\bottomrule
		\end{tabular}}
		\begin{tablenotes}
		\footnotesize
		\item[1] $d_\mathrm{B-0}$: the distance from vehicle $0$ to point B; $d_\mathrm{B-C}$: 
		the length of the road segment $B \rightarrow C$.
		\end{tablenotes}
		\end{threeparttable}
	\end{center}
	\vspace{-4mm}
\end{table}

\subsubsection{Ring-road experiment} 

Motivated by the seminal field experiments~\cite{sugiyama2008traffic,stern2018dissipation}, we further consider a single-lane circular track scenario of circumference $6.77\,\mathrm{m}$. As shown in Fig.~\ref{Fig:Overview_Ring}, a fleet of $9$ mini-vehicles are involved in this experiment, driving in the counter-clockwise direction ($E \rightarrow F\rightarrow G\rightarrow H\rightarrow E$). Compared with the straight-road experiment, there is no head vehicle that leads the whole fleet and each vehicle needs to follow its predecessor. Additionally, there is no external disturbance incorporated in the experiment. We aim to reproduce the traffic wave phenomenon without bottlenecks~\cite{sugiyama2008traffic}. 
Similarly to the experimental setup in~\cite{stern2018dissipation}, four phases are included in this experiment:
\begin{enumerate}[label=(\alph*)]
    \item In the beginning, all the vehicles are uniformly distributed on the ring-road. 
    \item Initially $t_1 = 0$, all the vehicles are HDVs (utilizing the OVM model) and they start to drive from idle velocity $0.05 \, \mathrm{m/s}$. 
    \item At some time point $t_2$ ($t_2 > t_1$), one vehicle is switched to CAV under \method{DeeP-LCC} control. \item At a later time point $t_3$ ($t_3 > t_2$), the CAV is switched back to HDV (utilizing the OVM model). All the vehicles continue to drive until the end of the experiment.    
\end{enumerate}

\section{Implementation of \method{DeeP-LCC} for \\ Mixed Traffic Flow}
\label{Sec:4}

This section presents the implementation methodology of \method{DeeP-LCC} in real experiments.

\subsection{Detailed Design of \method{DeeP-LCC}}

We first present some details in the \method{DeeP-LCC} formulation~\eqref{Eq:AdaptedDeePCforNonlinearSystem}. A quadratic form of the cost function $J(y,u)$ is under consideration
\begin{equation} \label{Eq:CostDefinition}
J(y,u) = \sum\limits_{k=t}^{t+N-1}\left( \left\|y(k)\right\|_{Q}^{2}+\left\|u(k)\right\|_{R}^{2}\right), 
\end{equation}
where $
Q=\diag(Q_v,Q_s)
$
with $Q_v = \diag(w_v,\ldots,w_v) \in \mathbb{R}^{n\times n}$, $Q_s = \diag(w_s,\ldots,w_s) \in \mathbb{R}^{m\times m}$ and $R = \diag(w_u,\ldots,w_u) \in \mathbb{R}^{m\times m} $. The specific weight coefficients $w_v,w_s,w_u$ represent the penalty for the velocity errors of all the vehicles, the spacing errors of all the CAVs, and the control inputs of the CAVs, respectively.

To guarantee driving safety and avoid control saturation, the input/output constraints are designed as follows. Both a lower bound and an upper bound are imposed on the spacing errors of the CAVs, in the sense that not only rear-end collisions for the CAVs can be avoided, but also the CAVs tend not to leave an extremely large inter-vehicle distance, which might cause other vehicles to cut in. Precisely, we have
\begin{equation} \label{Eq:Constraint_Spacing}
	\tilde{s}_\mathrm{min} \leq \tilde{s}_{i} \leq \tilde{s}_\mathrm{max} , \;\; i\in S,
\end{equation}
where $\tilde{s}_\mathrm{min},\tilde{s}_\mathrm{max}$ denote the minimum and maximum spacing error, respectively. Given the definition of system output~\eqref{Eq:SystemOutput}, the output constraint $y\in \mathcal{Y}$ in the original \method{DeeP-LCC} formulation~\eqref{Eq:AdaptedDeePCforNonlinearSystem} is then specified as 
\begin{equation} \label{Eq:SafetyConstraint}
	\tilde{s}_\mathrm{min} \leq I_{N} \otimes \begin{bmatrix}
	\mathbb{0}_{m \times n} & I_m
	\end{bmatrix} y \leq \tilde{s}_\mathrm{max}.
\end{equation}
Considering the practical actuation limit of individual vehicles, the input constraint $u\in \mathcal{U}$ in~\eqref{Eq:AdaptedDeePCforNonlinearSystem} is formulated as
\begin{equation} \label{Eq:AccelerationConstraint}
		a_\mathrm{min} \leq u \leq a_\mathrm{max},
\end{equation} 
where $	a_\mathrm{min}$ and $a_\mathrm{max}$ denote the minimum and the maximum acceleration, respectively. 

For the estimation $\hat{\epsilon}$ of the external input $\epsilon$, \ie, the velocity error of the head vehicle from the equilibrium velocity $v_0 - v^*$, we assume that
\begin{equation} \label{Eq:FutureExternalInput}
\hat{\epsilon} = \mathbb{0}_N,
\end{equation}
where $\mathbb{0}_N$ denotes a $N \times 1$ vector of all zeros. This assumption is valid since the head vehicle always attempts to maintain its current equilibrium velocity. With an appropriate estimation of the equilibrium velocity, which can be deduced from historic trajectories~\cite{jin2018experimental,jin2018connected}, this assumption~\eqref{Eq:FutureExternalInput} has been confirmed in traffic simulations~\cite{wang2022deeplcc} to be able to yield satisfactory control performance. We will provide further details on the estimation of the real-time equilibrium velocity in Section~\ref{Sec:OnlineControl}.

With this detailed design, the \method{DeeP-LCC} problem~\eqref{Eq:AdaptedDeePCforNonlinearSystem} can be specified as 
\begin{equation} \label{Eq:DeeP-LCC_Experiment}
 \begin{aligned}
 \min_{g,u,y,\sigma_y} \; &\sum\limits_{k=t}^{t+N-1}\!\!\left( \left\|y(k)\right\|_{Q}^{2}+\left\|u(k)\right\|_{R}^{2}\right)\!+\!\lambda_g \left\|g\right\|_2^2\!+\!\lambda_y \left\|\sigma_y\right\|_2^2\\
 \st \; & \begin{bmatrix}
 U_\mathrm{p} \\ E_\mathrm{p}\\Y_\mathrm{p} \\ U_\mathrm{f} \\ E_\mathrm{f}\\ Y_\mathrm{f}
 \end{bmatrix}g=
 \begin{bmatrix}
 u_\mathrm{ini} \\ \epsilon_\mathrm{ini}\\ y_\mathrm{ini} \\ u \\\epsilon \\ y
 \end{bmatrix}+\begin{bmatrix}
 0\\0\\ \sigma_y \\0 \\0 \\0
 \end{bmatrix},\\ &\epsilon = \mathbb{0}_N,\\
 & a_\mathrm{min} \leq u \leq a_\mathrm{max}, \\
 &\tilde{s}_\mathrm{min} \leq I_{N} \otimes \begin{bmatrix}
	\mathbb{0}_{m \times n} & I_m
	\end{bmatrix} y \leq \tilde{s}_\mathrm{max}.
 \end{aligned}
 \end{equation}
This is the final optimization control problem solved at each time step in the experiments. Problem~\eqref{Eq:DeeP-LCC_Experiment} can be further reformulated as a quadratic program (QP), and be solved efficiently via commercial solvers. The MATLAB QP solver \method{quadprog} is employed in the experiments.

\subsection{Implementation of \method{DeeP-LCC}}
\label{Sec:Implementation}

We now present the implementation of \method{DeeP-LCC}, which consists of two parts: offline data collection and online predictive control.

\subsubsection{Offline Data Collection} 

Before presenting the data collection process, we first need to clarify the vehicles that are incorporated into the \method{DeeP-LCC} formulation. In practice, it is not necessary to collect the driving data of all the vehicles in traffic flow  for CAV controller design. 
\begin{itemize}
    \item In the straight-road experiment shown in Fig.~\ref{Fig:Overview_Straight}, we have $6$ vehicles in total, and all the vehicles are involved in \method{DeeP-LCC}. Vehicle $0$ is literally the first vehicle of the fleet, and also serves as the ``head vehicle" in \method{DeeP-LCC} (see Fig.~\ref{Fig:SystemSchematic}). Different cases of the CAV number and locations, \ie, the choice of $S$, will be under consideration.
    \item In the ring-road experiment shown in Fig.~\ref{Fig:Overview_Ring}, there exist $9$ vehicles in the traffic system, but only $5$ of them, vehicles $3-7$, are incorporated into \method{DeeP-LCC}. Particularly, vehicle $3$ is regarded as the head vehicle in \method{DeeP-LCC} formulation, and vehicle $5$ will be switched to CAV at certain time period.
\end{itemize}

For each choice of $S$, \ie, the CAV number and locations, independent experiments are carried out for data collection in \method{DeeP-LCC}. For the HDVs, \ie, vehicle $i$ with $i \in \Omega \backslash S$, they have a fixed but unknown car-following dynamics~\eqref{Eq:OVMmodel}. For the CAVs ($i \in S$), it is required to have a prior controller with acceptable performance in data collection. One natural way is to adopt the HDVs' car-following model as the CAVs' controller, where random disturbances are added on the control input to satisfy the persistent excitation requirement in Assumption~\ref{Assumption:PersistentExcitation}. Accordingly, we design the following CAVs' control input for data collection based on the OVM dynamics~\eqref{Eq:OVMmodel} 
\begin{equation} \label{Eq:Collection_ControlInput}
   u_i(t) = \alpha\left(v_{\mathrm{des}}\left(s_i(t)\right)-v_i(t)\right)+\beta\dot{s}_i(t) + \mathbb{U}[-\delta_u,\delta_u], 
\end{equation}
where $\mathbb{U}[\cdot,\cdot]$ denotes the uniform distribution and $\delta_u>0$ denotes the perturbation upper bound for control input. Note that although we utilize a similar OVM model to control the CAVs in data collection, the CAVs have no knowledge of the driving behavior of the surrounding HDVs.

Additionally, the external input in \method{DeeP-LCC}, \ie, the velocity of the head vehicle, should also be carefully designed, in order to guarantee the persistent excitation of combined input (consisting of control input and external input), as required in Assumption~\ref{Assumption:PersistentExcitation}. Precisely,
\begin{itemize}
    \item For the straight-road experiment, the velocity of the head vehicle is designed as 
    \begin{equation} \label{Eq:Collection_ExternalInput_Straight}
   \epsilon(t) = \mathbb{U}[-\delta_\epsilon,\delta_\epsilon], 
\end{equation}
    where $\delta_\epsilon>0$ denotes the perturbation upper bound for the velocity of the head vehicle. This design allows the head vehicle to maintain the prescribed constant velocity $v_\mathrm{c}$, while also suffering from random perturbations.
    \item For the ring-road experiment, the head vehicle, \ie, vehicle $3$, needs to follow its own predecessor. Motivated by the control strategy in~\cite{cui2017stabilizing}, its acceleration is designed as
    \begin{equation} \label{Eq:Collection_ExternalInput_Ring}
    \begin{aligned}
       \epsilon(t) = & \alpha\left(v_{\mathrm{des}}\left(s_i(t)\right)-v_i(t)\right)+\beta\dot{s}_i(t) \\ & - k_r(v_i(t)-v_r) + \mathbb{U}[-\delta_u,\delta_u].
    \end{aligned}
\end{equation}
    In~\eqref{Eq:Collection_ExternalInput_Ring}, the original OVM dynamics is utilized as~\eqref{Eq:Collection_ControlInput} in CAV's control input for data collection. Moreover, a feedback term $k_r(v_i(t)-v_r)$ is also added, enabling the head vehicle to maintain a reference equilibrium velocity $v_r = 0.25 \, \mathrm{m/s}$ in data collection; otherwise, stop-and-go waves might quickly occur and the collected data is far from the equilibrium state.
\end{itemize}

\noindent Under the design of~\eqref{Eq:Collection_ControlInput} combined with~\eqref{Eq:Collection_ExternalInput_Straight} or~\eqref{Eq:Collection_ExternalInput_Ring}, given sufficient data, the persistent excitation requirement in Assumption~\ref{Assumption:PersistentExcitation} is often satisfied, as verified in our experiments. 

Then, we carry out independent experiments and collect the original measurable traffic data for \method{DeeP-LCC}, including the control input sequence $u^\mathrm{d}$, the external input sequence $\epsilon^\mathrm{d}$, and raw output sequence $y_\mathrm{raw}^\mathrm{d}$. Particularly, instead of the output $y(t)$ defined in~\eqref{Eq:SystemOutput}, we consider the raw output at time $t$, defined as
\begin{equation} \label{Eq:RawOutput}
    y_\mathrm{raw}(t) \!=\!\begin{bmatrix}v_{1}(t),v_{2}(t),\ldots,{v}_{n}(t),{s}_{i_1}(t),{s}_{i_2}(t),\ldots,{s}_{i_m}(t)\end{bmatrix}^{\top} ,
\end{equation}
where $y_\mathrm{raw}(t) \in \mathbb{R}^{n+m}$ contains no information of the equilibrium traffic state, but consists of the original data of vehicle velocity and spacing. This raw output signal and the defined output signal  have the following relationship
\begin{equation} 
    y(t) = y_\mathrm{raw}(t) - [
    \overbrace{v^*,\ldots,v^*}^{n},\overbrace{s^*,\ldots,s^*}^{m}
    ]^\top.
\end{equation}
By design, we have prior knowledge of equilibrium velocity in offline data collection, which is $v^*=v_\mathrm{c}$ in straight road experiments and $v^*=v_r$ in ring road experiments. The equilibrium spacing $s^*$ is also a designed value. Therefore, we can obtain the pre-collected data sequences $u^\mathrm{d},\epsilon^\mathrm{d},y^\mathrm{d}$, and then construct data Hankel matrices by~\eqref{Eq:DataHankel}.

\subsubsection{Online Predictive Control}
\label{Sec:OnlineControl}

The online control process is sequentially divided into two steps: 1) Past data collection and equilibrium estimation, and 2) control computation and implementation. This process is executed in a receding horizon manner.

\vspace{1mm}
\emph{(a) Past data collection and equilibrium estimation} 
\vspace{1mm}

During control process, the following raw past data before current time $t$ should be collected:
\begin{equation} \label{Eq:RawPastData}
    \begin{cases}
    u(t-T_\mathrm{ini}),u(t-T_\mathrm{ini}+1),\ldots,u(t-1);\\
    v_0(t-T_\mathrm{ini}),v_0(t-T_\mathrm{ini}+1),\ldots,v_0(t-1);\\
    y_\mathrm{raw}(t-T_\mathrm{ini}),y_\mathrm{raw}(t-T_\mathrm{ini}+1),\ldots,y_\mathrm{raw}(t-1),\\
    \end{cases}
\end{equation}
which correspond to control input, head vehicle velocity, and raw output, respectively. The control input sequence can be directly utilized to construct $u_\mathrm{ini}$ by~\eqref{Eq:PastDataDefinition}, but to obtain $\epsilon_\mathrm{ini},y_\mathrm{ini}  $, the current equilibrium state needs to estimated.

\begin{algorithm}[t]
	\caption{Implementation of \method{DeeP-LCC}}
	\label{Alg:DeeP-LCC}
	\begin{algorithmic}[1]
		\Require
		Pre-collected traffic data Hankel matrices $U_{\mathrm{p}},U_{\mathrm{f}},$ $E_{\mathrm{p}},E_{\mathrm{f}},Y_{\mathrm{p}},Y_{\mathrm{f}}$, initial time $t_0$, terminal time $t_f$;
		\State Initialize past traffic data $(u_{\mathrm{ini}},\epsilon_{\mathrm{ini}},y_{\mathrm{ini}})$ before $t_0$;
		\While{$t_0 \leq t \leq t_f$}
		\State Collect raw past data in~\eqref{Eq:RawPastData};
		\State Estimate equilibrium velocity $v^*$ by~\eqref{Eq:EquilibriumVelocity} and design equilibrium spacing $s^*$ by~\eqref{Eq:EquilibriumSpacing};
		\State Update past traffic data $(u_{\mathrm{ini}},\epsilon_{\mathrm{ini}},y_{\mathrm{ini}})$;
		\State Solve~\eqref{Eq:DeeP-LCC_Experiment} for optimal control input $u_\mathrm{opt}=\col\left(u_\mathrm{opt}(t),u_\mathrm{opt}(t+1),\ldots,u_\mathrm{opt}(t+N-1)\right)$;
		\State Apply the input $u(t) \leftarrow u_\mathrm{opt}(t), u(t+1) \leftarrow u_\mathrm{opt}(t+1),\ldots,u(t+N_\mathrm{c}-1) \leftarrow u_\mathrm{opt}(t+N_\mathrm{c}-1)$ to the CAVs;
		\State $t \leftarrow t+N_\mathrm{c}$;
		\EndWhile
	\end{algorithmic}
\end{algorithm}

Particularly,  the equilibrium velocity $v^* $ of traffic flow needs to be estimated and updated, while the equilibrium spacing $s^*$ for the CAVs needs to be designed. At the time step $t$, we utilize a natural approach for equilibrium velocity estimation, given by
\begin{equation} \label{Eq:EquilibriumVelocity}
    v^*= \frac{1}{T_\mathrm{ini}}\sum_{t-T_\mathrm{ini}}^{t-1}v_0(t),
\end{equation}
which is obtained by averaging the past velocity of the head vehicle from time $t-T_\mathrm{ini}$ to time $t-1$. Note that the time horizon $T_\mathrm{ini}$ is consistent with the past time horizon in \method{DeeP-LCC}, and thus we can utilize the same collected past data~\eqref{Eq:RawPastData} in the control process for both equilibrium estimation and \method{DeeP-LCC} formulation~\eqref{Eq:DeeP-LCC_Experiment}. For CAVs' equilibrium spacing, we consider a velocity-dependent spacing policy, identical to the human driving policy in OVM~\eqref{Eq:OVMDesiredVelocity} which describes the relationship between current spacing and desired velocity. Precisely, we have
\begin{equation} \label{Eq:EquilibriumSpacing}
    s^*= \arccos\left(1-2\frac{v^*}{v_{\mathrm{max}}}\right)\cdot \frac{s_\mathrm{go}-s_\mathrm{st}}{\pi} + s_{\mathrm{st}},
\end{equation}
which is the inverse function of~\eqref{Eq:OVMDesiredVelocity} given $0 \leq v^* \leq v_\mathrm{max}$. Note that although we design a human-type spacing policy for the CAVs, the CAVs have no knowledge of the car-following dynamics of the surrounding HDVs, as well as their individual spacing policy.

With the equilibrium information $(v^*,s^*)$ available, the past sequences of external input $\epsilon_\mathrm{ini}$ and traffic output $y_\mathrm{ini}$ before time $t$ can be calculated based on the raw data in~\eqref{Eq:RawPastData}. 

\vspace{1mm}
\emph{(b) Control computation and implementation} 
\vspace{1mm}

The \method{DeeP-LCC} problem~\eqref{Eq:DeeP-LCC_Experiment} is numerically solved and the obtained optimal control inputs are applied. At the time step $t$, solving~\eqref{Eq:DeeP-LCC_Experiment} provides the optimal control sequence in the future time horizon $N$: 
\begin{equation}
    u_\mathrm{opt}=\col\left(u_\mathrm{opt}(t),u_\mathrm{opt}(t+1),\ldots,u_\mathrm{opt}(t+N-1)\right).
\end{equation}
The first $N_\mathrm{c}$ ($N_\mathrm{c}\in\mathbb{N}, N_c \leq N$) steps of control inputs are applied into the system, and the aforementioned process is repeated in a receding horizon manner after setting time $t$ to time $t+N_\mathrm{c}$. The implementation procedures are shown in Algorithm~\ref{Alg:DeeP-LCC}. The code for \method{DeeP-LCC} is available at {\small \url{https://github.com/soc-ucsd/DeeP-LCC}}.

\begin{remark}[Time-varying equilibrium states]
Compared with standard DeePC~\cite{coulson2019data}, \method{DeeP-LCC} includes an equilibrium estimation process for implementation, since practical traffic flow has time-varying equilibrium states. In our experiments, the offline data collection is carried out around one prescribed equilibrium state (see Fig.~\ref{Fig:SystemSchematic}). Since the system output~\eqref{Eq:SystemOutput} is defined as error signals from equilibrium, we assume that the error dynamics of mixed traffic around different equilibrium states are similar to each other, and they are already captured in the data Hankel matrices~\eqref{Eq:DataHankel}. Still, we have observed that such a simple implementation strategy has allowed \method{DeeP-LCC} for effectively mitigating traffic perturbations in the experiments. Note that Algorithm~\ref{Alg:DeeP-LCC}  may have a comprised optimization performance, due to the inconsistency between the traffic equilibrium when collecting data and the current traffic condition when applying data-driven predictive control. One future direction is thus to collect traffic data from different equilibrium states and adaptively choose appropriate data in the online predictive control process.
\end{remark}

\begin{table}[tb]
\footnotesize
	\begin{center}
		\caption{ Parameter Setup for \method{DeeP-LCC}}
		\vspace{-2mm}
		\label{Tb:DeePLCC}
		\begin{threeparttable}
		\setlength{\tabcolsep}{2mm}{
		\begin{tabular}{ccc}
		\toprule
			Symbol & Meaning & Value \\\hline
			$\delta_\epsilon $ & Perturbation in~\eqref{Eq:Collection_ExternalInput_Straight} for data collection & $0.05$  \\
			$\delta_u $ & Perturbation in~\eqref{Eq:Collection_ControlInput},~\eqref{Eq:Collection_ExternalInput_Ring} for data collection & $0.2$\\
			$k_r$ & Feedback gain in~\eqref{Eq:Collection_ExternalInput_Ring} for data collection & $8$\\
			$T$ & Length for pre-collected data  & $1500$ \\
			$N$ & Length for future data horizon  & $50$ \\
			$T_{\mathrm{ini}}$ & Length for past data horizon & $20$ \\
			$N_\mathrm{c}$ & Control horizon for predicted input  & $10$ \\
			$\tilde{s}_{\max}$ & Upper bound of spacing error  & $1.2$ \\
			$\tilde{s}_{\min}$ & Lower bound of spacing error  & $-0.4$ \\
			$a_{\max}$ & Upper bound of acceleration  & $0.4$ \\
			$a_{\min}$ & Lower bound of acceleration  & $-0.4$ \\
			$w_v$ & Weight coefficient for velocity error  & $5$ \\
			$w_s$ & Weight coefficient for spacing error  & $40/k$ \\
			$w_u$ & Weight coefficient for control input  & $2/k$  \\
			$\lambda_g$ & Regularized coefficient for $g$  & $10$ \\
			$\lambda_y$ & Regularized coefficient for $\sigma_y$  & $10^5$ \\
			\bottomrule
		\end{tabular}}
		\begin{tablenotes}
		\footnotesize
		\item[1] $k$ represents the CAV number, which could be $1$ or $2$ in the experiments.
		\end{tablenotes}
		\end{threeparttable}
	\end{center}
	\vspace{-5mm}
\end{table}

\section{Experimental Results}
\label{Sec:5}

\begin{figure*}[htb!]
	\vspace{2mm}
	\centering
	\subfigure[All the vehicles are HDVs]
	{
	\includegraphics[width=8cm]{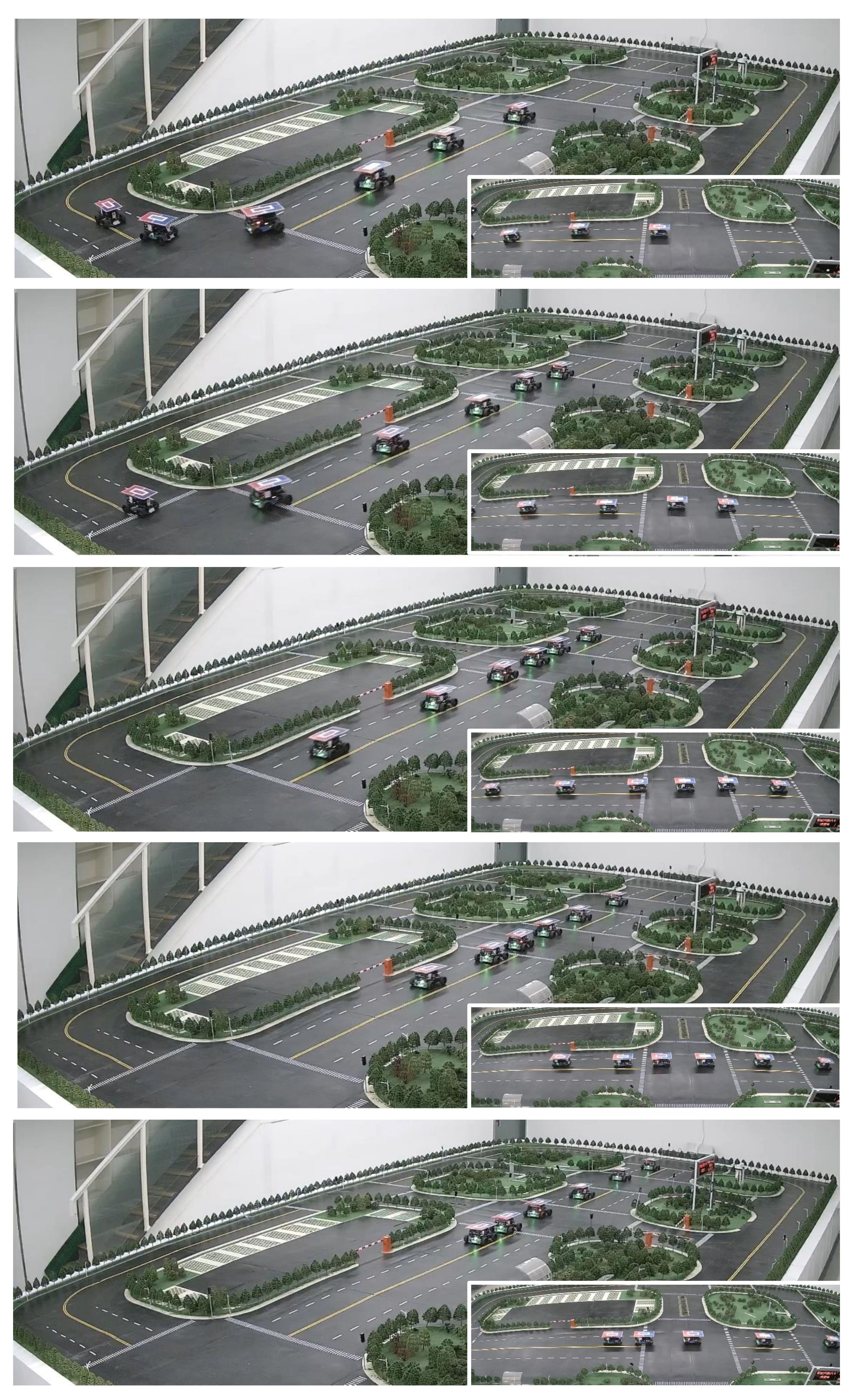}
	\label{Fig:Snapshot_000000}
	}
	\subfigure[\method{DeeP-LCC} with $S=\{2\}$]
	{
	\includegraphics[width=8cm]{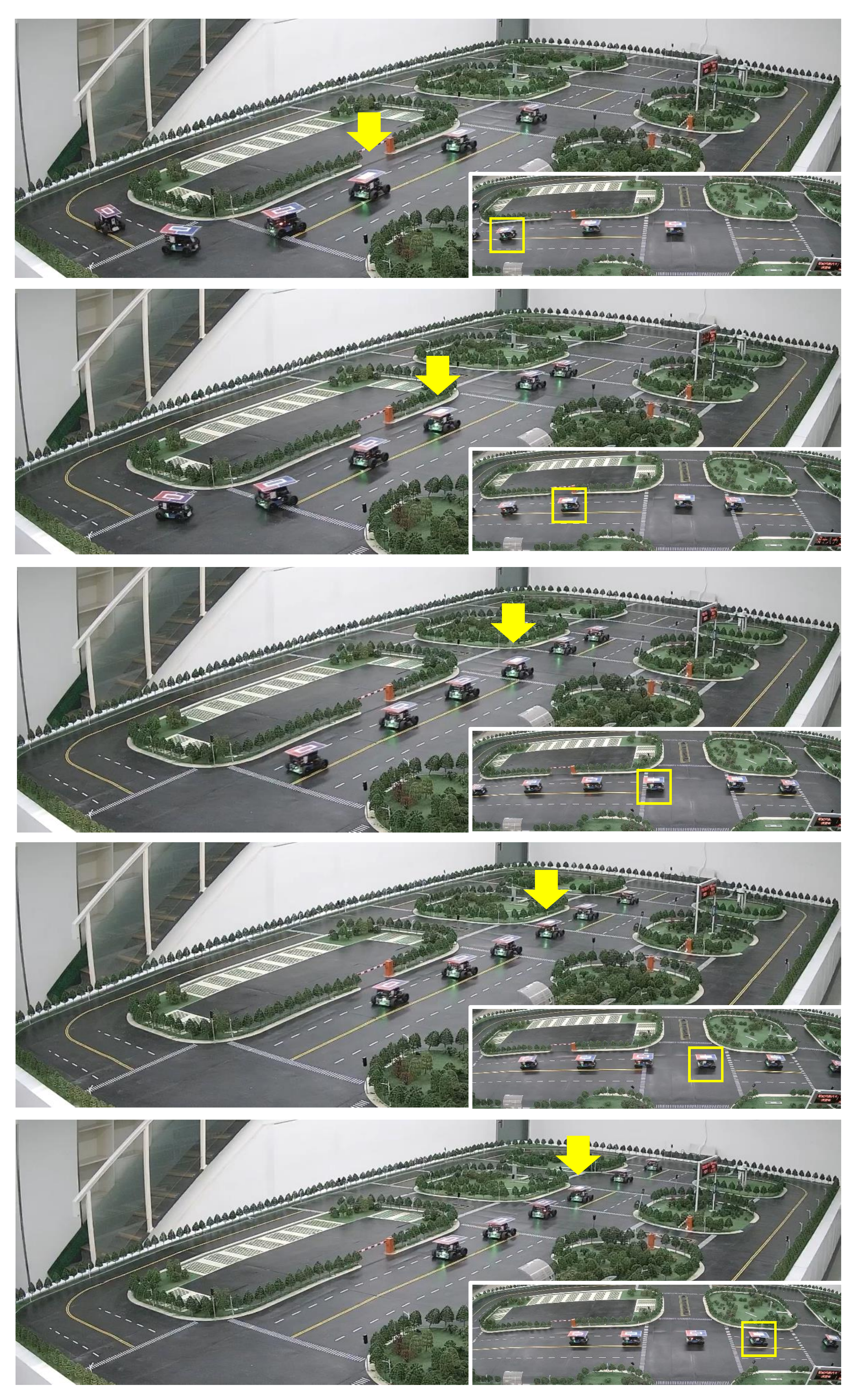}
	\label{Fig:Snapshot_001000}
	}
	\vspace{-2mm}
	\caption{Snapshots of the straight-road experiments under different cases of mixed traffic. The vehicle fleet is running on the $B \rightarrow C$ road segment, as shown in Fig.~\ref{Fig:Overview_Straight}. Different camera angles are incorporated to provide a full view of the experimental process. The complete video can be found in \url{https://www.youtube.com/watch?v=ZZ2cWhapqpc}. (a) All the vehicles are HDVs. (b) The vehicle indexed as $2$ (see the vehicle below the yellow angle, or the vehicle highlighted in the yellow box) is CAV under \method{DeeP-LCC} control.  }
	\label{Fig:Snapshot}
	\vspace{-3mm}
\end{figure*}

This section presents the experimental results for validation of \method{DeeP-LCC}. The parameters of the OVM model for HDVs are listed in Table~\ref{Tb:OVM}, and the \method{DeeP-LCC} parameter setup is listed in Table~\ref{Tb:DeePLCC}. The sampling time interval $\Delta t$ for \method{DeeP-LCC} offline data collection is chosen to be $50\,\mathrm{ms}$, and the offline data are all collected around an velocity of $0.3 \, \mathrm{m/s}$.

\subsection{Straight-road Experiments}

In the straight-road experiments, recall that the head vehicle drives according to the velocity trajectory in Table~\ref{Tb:HeadVehicleTrajectory}. For the following five vehicles, we consider various  mixed traffic cases depending on the number and location of the CAV(s), \ie, the choice of $S$. Particularly, we consider five cases: $S=\varnothing,\{1\},\{2\},\{1,3\},\{2,4\}$ (the case $S=\varnothing$ corresponds to the scenario where all the vehicles are HDVs).

\begin{figure*}[htb]
	\vspace{2mm}
	\centering
	\subfigure[All the vehicles are HDVs]
	{
	\includegraphics[width=8cm]{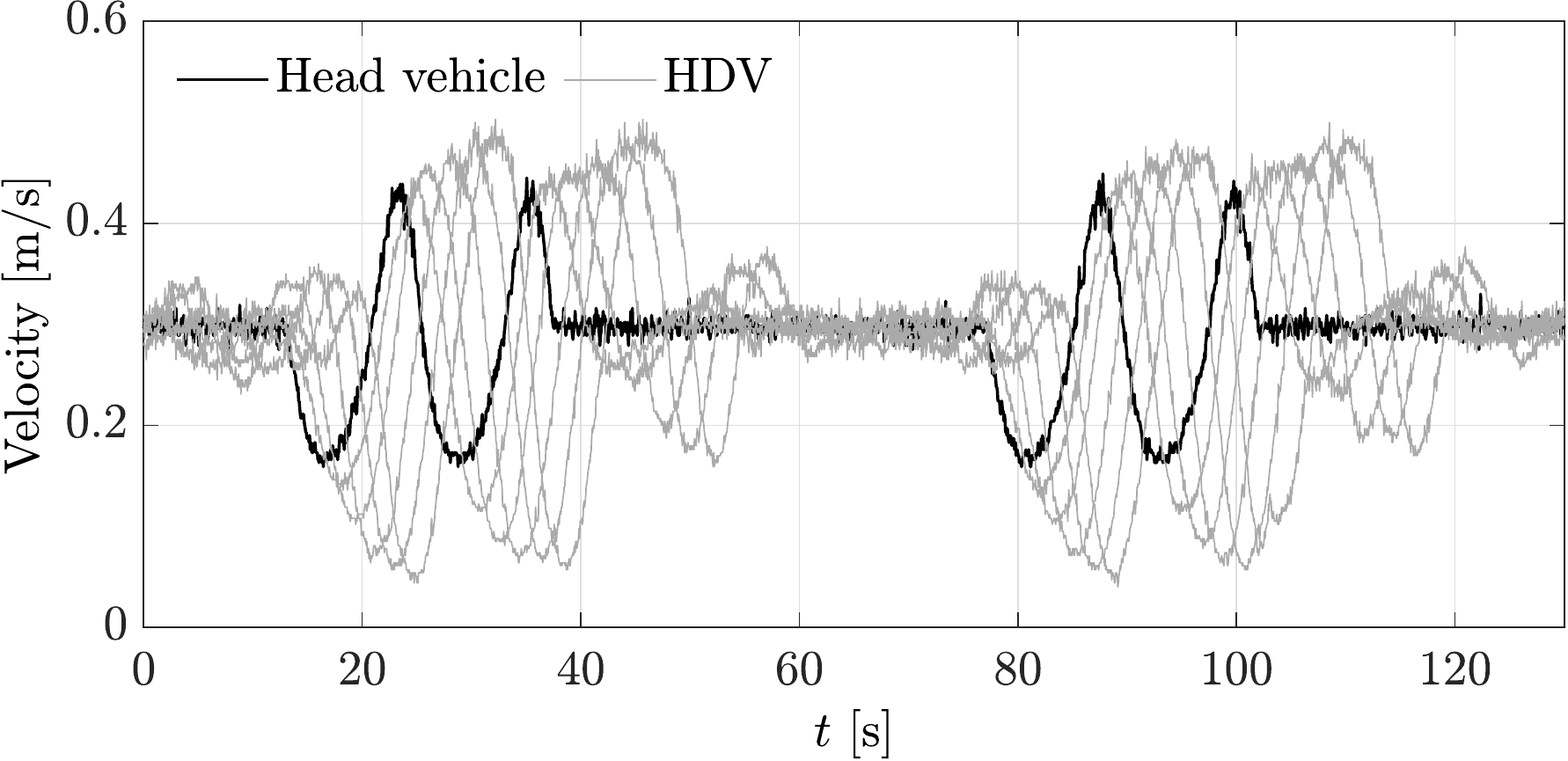}
	\label{Fig:Velocity_000000}
	}
	\\
	\subfigure[\method{DeeP-LCC} with $S=\{1\}$]
	{
	\includegraphics[width=8cm]{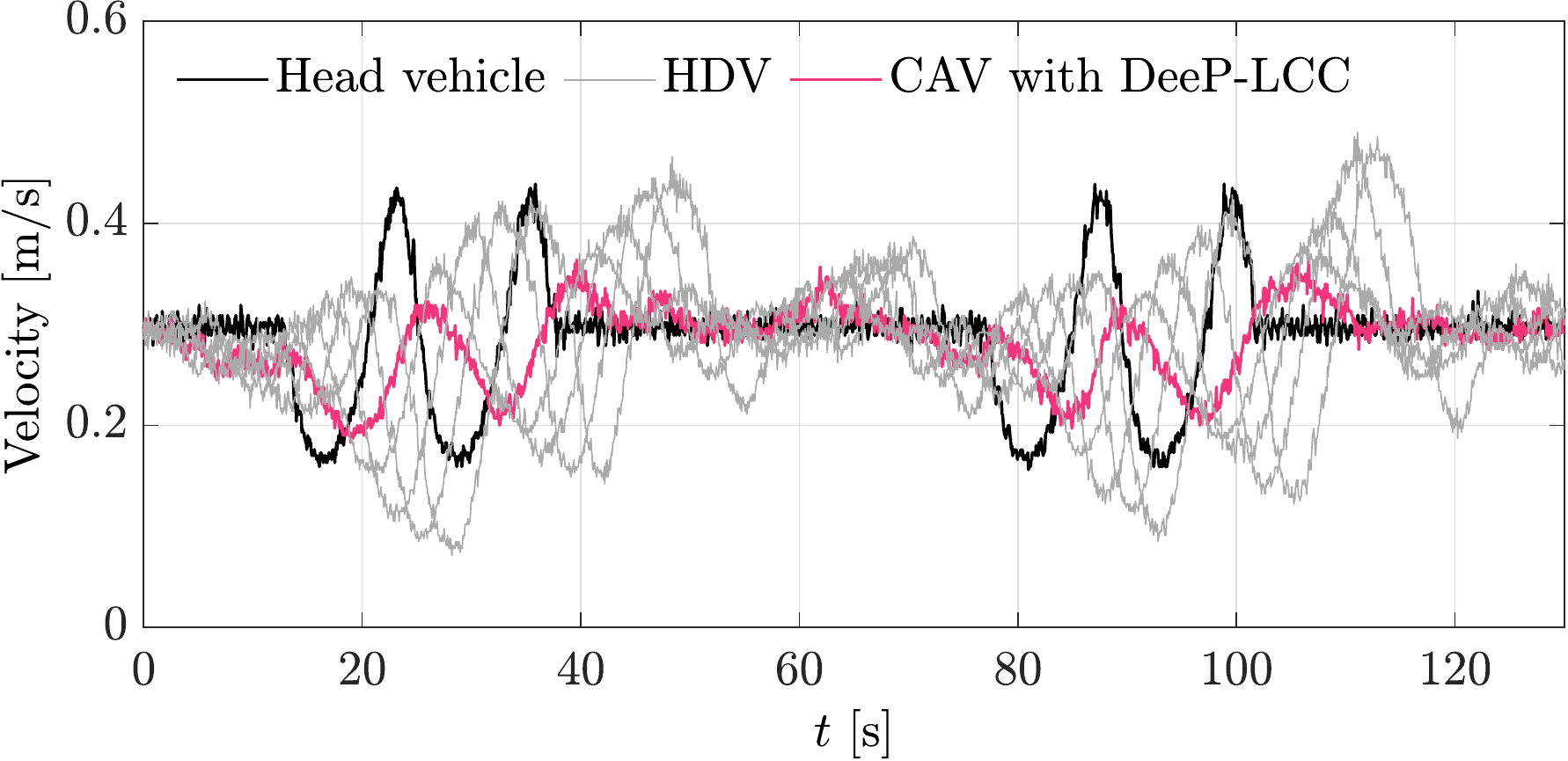}
	\label{Fig:Velocity_010000}
	} \hspace{5mm}
	\subfigure[\method{DeeP-LCC} with $S=\{2\}$]
	{
	\includegraphics[width=8cm]{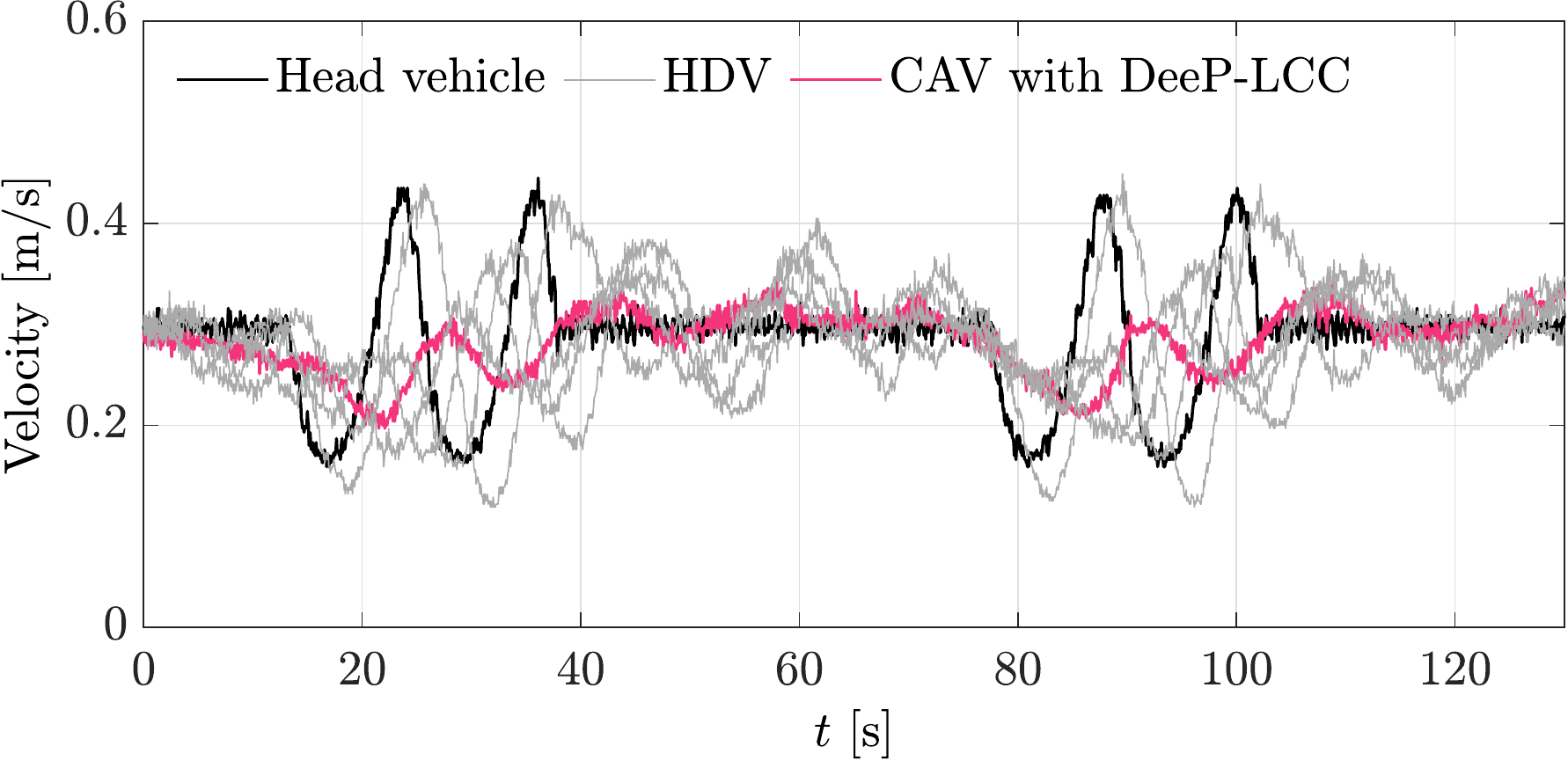}
	\label{Fig:Velocity_001000}
	}
	\subfigure[\method{DeeP-LCC} with $S=\{1,3\}$]
	{
	\includegraphics[width=8cm]{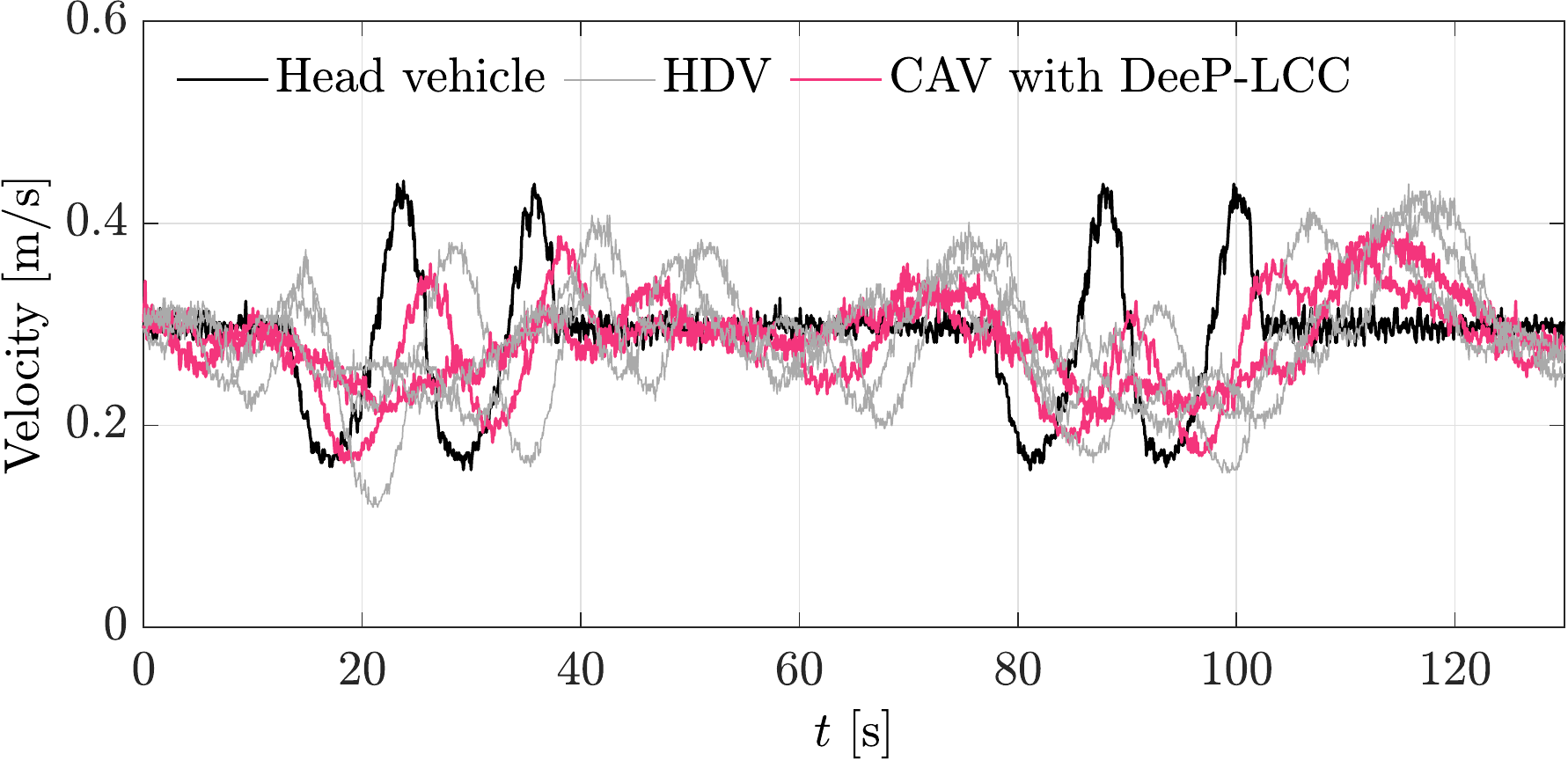}
	\label{Fig:Velocity_010100}
	} \hspace{5mm}
	\subfigure[\method{DeeP-LCC} with $S=\{2,4\}$]
	{
	\includegraphics[width=8cm]{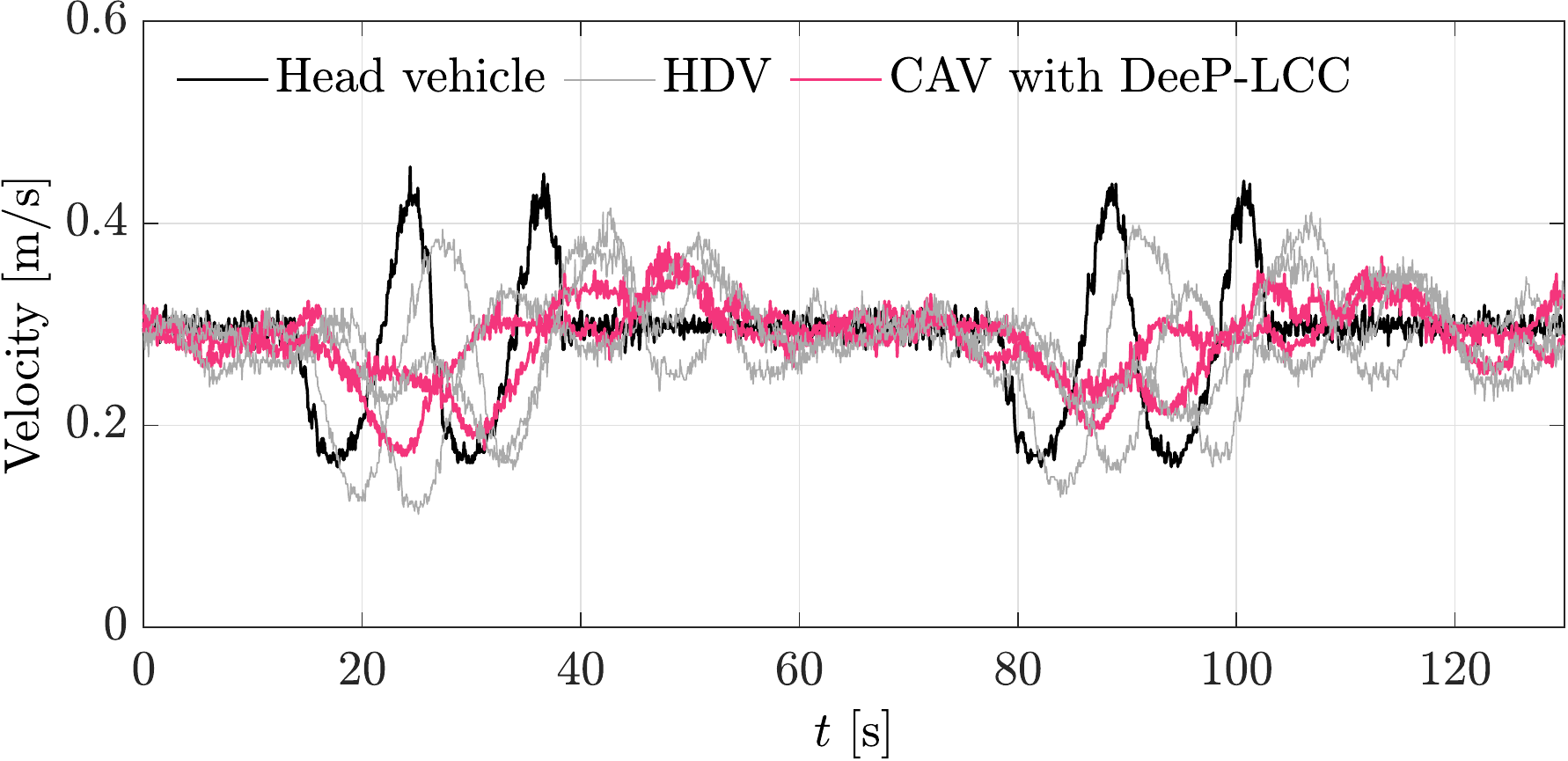}
	\label{Fig:Velocity_001010}
	}
	\vspace{-2mm}
	\caption{Velocity profiles of all the vehicles in the straight-road experiments under different cases of mixed traffic. The black profiles, gray profiles, and red profiles represent the velocity of the head vehicle, the HDVs and the CAV, respectively.}
	\label{Fig:Velocity_straight}
	\vspace{-4mm}
\end{figure*}

Fig.~\ref{Fig:Snapshot} demonstrates the snapshots of the recorded video from different viewing angles in the experimental platform. The velocity profiles in different cases of $S$ are shown in Fig.~\ref{Fig:Velocity_straight}, where the head vehicle is shown in black and the following vehicles are colored in gray for HDVs and red for CAVs. As clearly observed from Fig.~\ref{Fig:Snapshot_000000}, when all the vehicles are HDVs, the entire vehicle fleet is caught up in the traffic wave triggered by the external sinusoid perturbation in the $B \rightarrow C$ road segment. The inter-vehicle distances are rapidly changing, and the velocity profiles in Fig.~\ref{Fig:Velocity_000000} further show that the amplitude of the velocity oscillation is amplified against the traffic moving direction, from the head to the tail. Even after the vehicle fleet passes the $B \rightarrow C$ road segment, \ie, the external disturbance is no longer imposed, an apparent traffic wave still persists for a while, until all the HDVs finally return to the equilibrium. This experiment reproduces the traffic wave phenomenon when there exist some apparent wave triggers, such as road bottlenecks or lane changes.

Then, one or several vehicles in this experiment are changed from HDVs to CAVs with \method{DeeP-LCC}. Fig.~\ref{Fig:Snapshot_001000} demonstrates the snapshot when the second following vehicle is CAV, \ie, $S=\{2\}$. It is well-observed that the traffic instabilities are apparently reduced and the following vehicles keep a relatively stable inter-vehicle distance. As shown in Fig.~\ref{Fig:Velocity_001000} (the velocity profiles), the propagation of traffic waves occurring from the sinusoid perturbation is mitigated to a large extent. Similar results are observed in other choices of $S$, including the case with one CAV or two CAVs; see Fig.~\ref{Fig:Velocity_straight} for demonstrations. These results reveal the capability of \method{DeeP-LCC} in dissipating stop-and-go waves induced from external perturbations.

\begin{table}[tb]
	\begin{center}
		\caption{Reduction of ASVE from Equilibrium at straight-road Experiments}
		\vspace{-2mm}
		\label{Tb:ASVE}
		\begin{threeparttable}
		\setlength{\tabcolsep}{2mm}{
		\begin{tabular}{ccc}
		\toprule
			& ASVE Reduction from EE & ASVE Reduction from PE \\\hline
			$S=\{1\}$ & $6.09\%$ & $50.15\%$ \\
			$S=\{2\}$ & $8.99\%$ & $67.96\%$  \\
			$S=\{1,3\}$ & $9.40\%$ & $67.68\%$  \\
			$S=\{2,4\}$ & $9.54\%$ & $73.14\%$  \\
			\bottomrule
		\end{tabular}}
		\begin{tablenotes}
		\footnotesize
		\item[1] EE: estimated equilibrium, which is set as~\eqref{Eq:EquilibriumVelocity}.
		\item[2] PE: prescribed equilibrium, which is set as $v_\mathrm{c} = 0.3\,\mathrm{m/s}$.
		\end{tablenotes}
		\end{threeparttable}
	\end{center}
	\vspace{-5mm}
\end{table}

We further use an index of accumulated squared velocity error (ASVE)~\cite{wang2021leading,wang2022deeplcc} to quantify the traffic instabilities at different mixed traffic cases (\ie, different values of $S$), given by
$
    \mathrm{ASVE} = \sum_1^n\int_{t_0}^{t_f}(v_i(t)-v^*)^2dt,
$
which sums up the integrated deviation of each vehicle from the equilibrium velocity over time. In particular, we consider two kinds of equilibrium velocity: 1) an estimated equilibrium velocity from historic trajectory of the head vehicle, with the same definition of $v^*$ in~\eqref{Eq:EquilibriumVelocity}; 2) a prescribed equilibrium velocity from designed trajectory of the head vehicle, which is the same as the value of $v_\mathrm{c}$ in Table~\ref{Tb:HeadVehicleTrajectory}.  
The reduction rate of ASVE under different choices of $S$ compared with the case where all the vehicles are HDVs is listed in Table~\ref{Tb:ASVE}. In all cases, \method{DeeP-LCC} enables the CAV(s) to dampen traffic perturbations, with slight performance differences depending on the penetration rate and locations of CAVs. Note that we have considered an identical parameter setup in different mixed traffic cases $S$, and thus the performance might be compromised without parameter tuning at each case of $S$. Nevertheless, these results have already validated the effectiveness and applicability of \method{DeeP-LCC} in stabilizing traffic flow in different market penetration rates and CAV spatial locations.

\begin{figure*}[tb!]
	\vspace{2mm}
	\centering
	\subfigure[Snapshot at time $0 \, \mathrm{s}$]
	{
	\includegraphics[width=8cm]{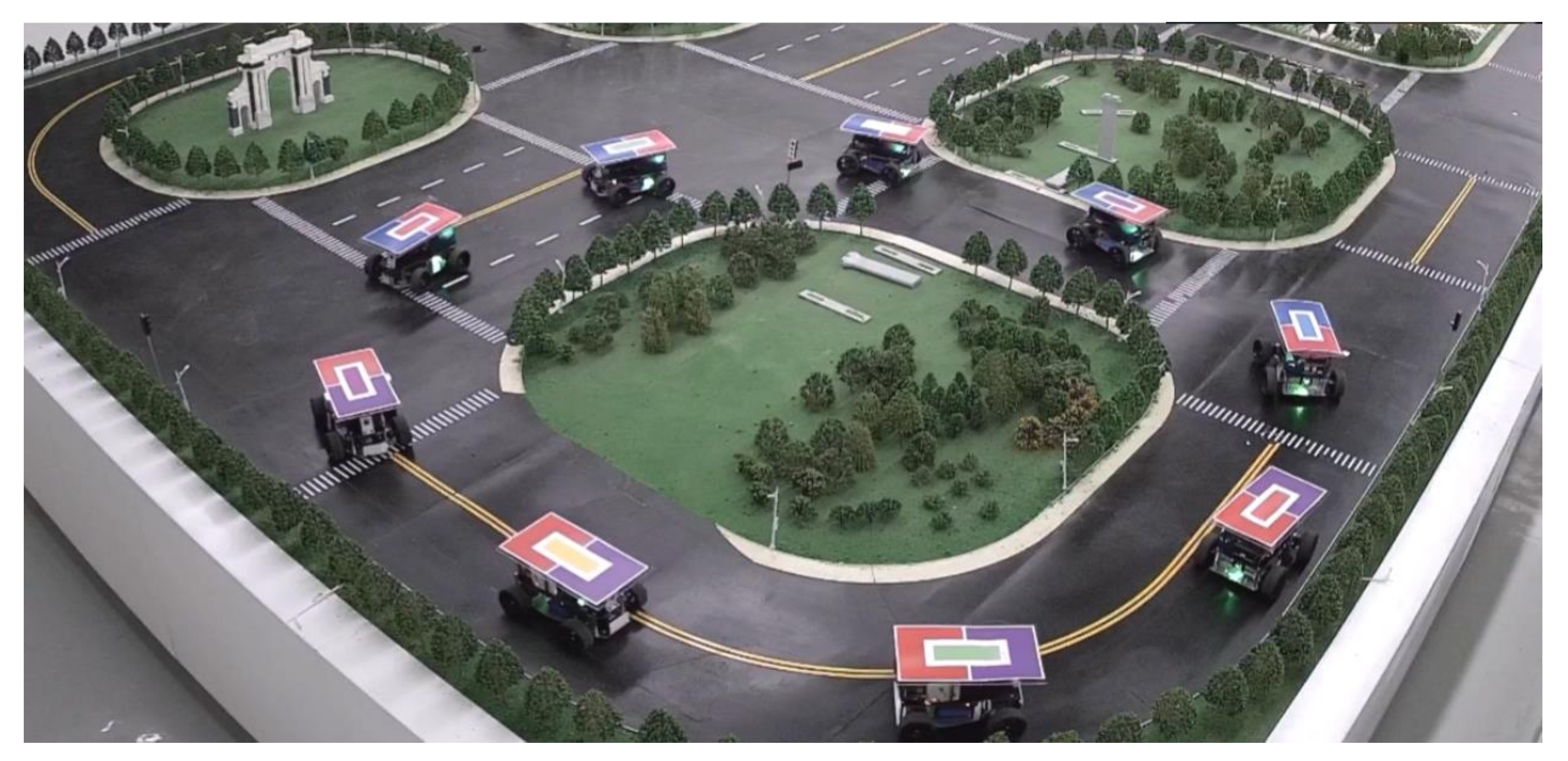}
	\label{Fig:Snapshot_ringroad_initial}
	}
	\subfigure[Snapshot at time $66.15 \, \mathrm{s}$ (all the vehicles are HDVs) ]
	{
	\includegraphics[width=8cm]{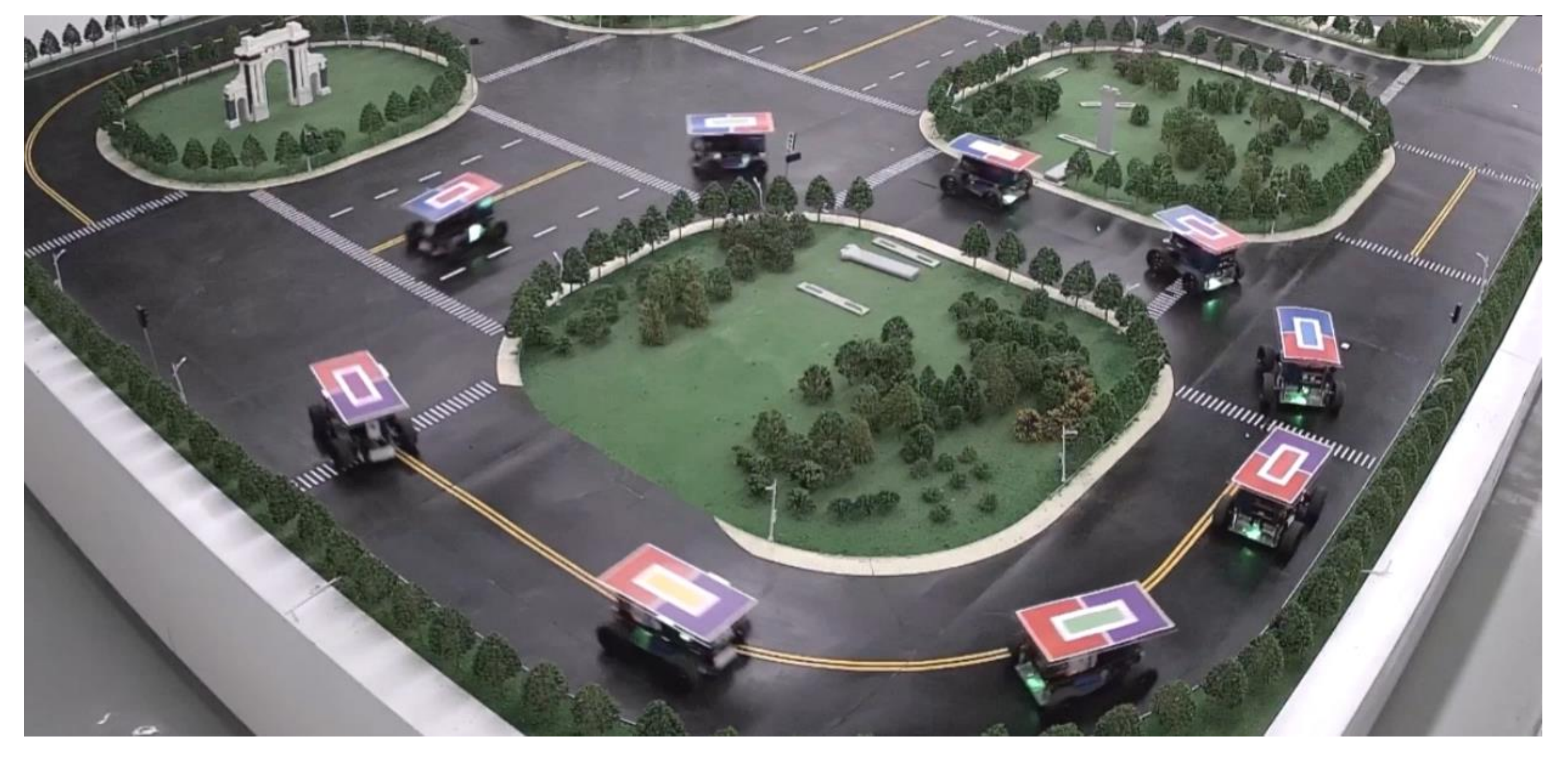}
	\label{Fig:Snapshot_ringroad_hdv}
	} \\
	\vspace{-2mm}
	\subfigure[ Snapshot at time $134.75 \, \mathrm{s}$ (\method{DeeP-LCC} is activated for one vehicle) ]
	{
	\includegraphics[width=8cm]{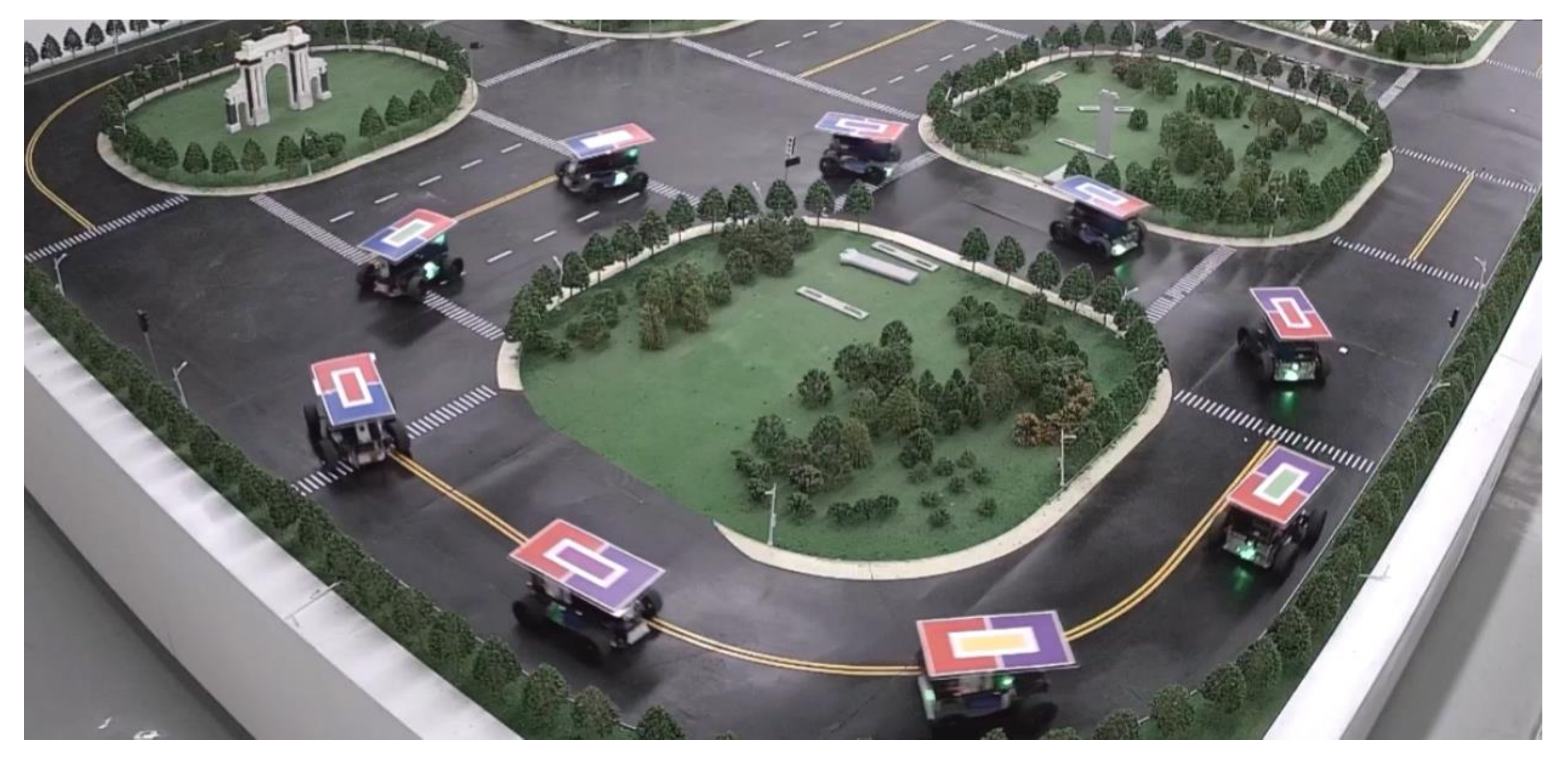}
	\label{Fig:Snapshot_ringroad_cav}
	}
	\subfigure[Snapshot at time $197.35 \, \mathrm{s}$ (all the vehicles are HDVs) ]
	{
	\includegraphics[width=8cm]{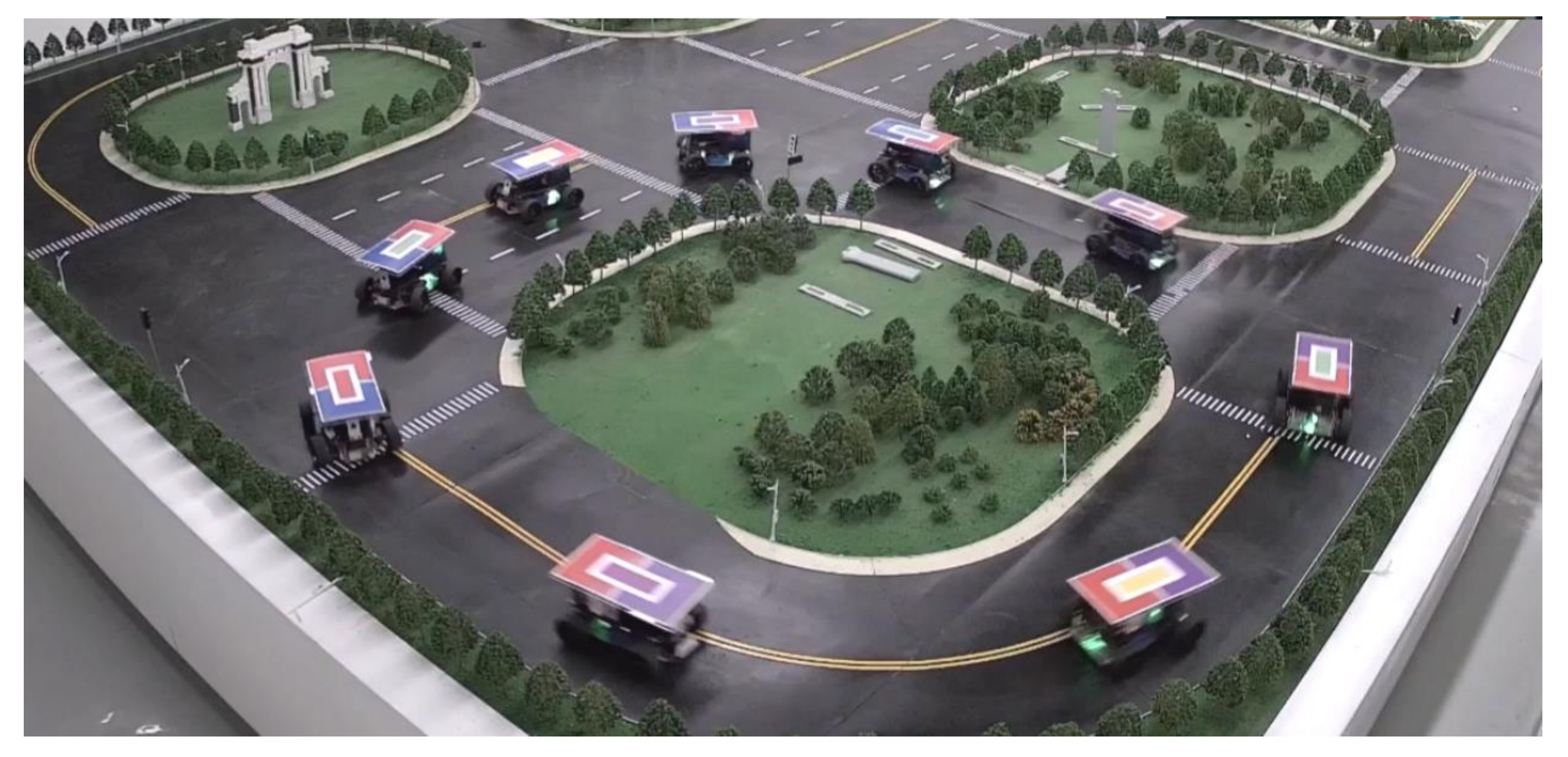}
	\label{Fig:Snapshot_ringroad_hdv2}
	}
	\vspace{-2mm}
	\caption{Snapshots of the ring-road experiments at different times. From time $0\,\mathrm{s}$ to time $68.85\,\mathrm{s}$, all the vehicles are HDVs under the OVM model. Since time $68.85\,\mathrm{s}$, \method{DeeP-LCC} is activated for vehicle $5$. After time $139.05\,\mathrm{s}$, \method{DeeP-LCC} is deactivated and all the vehicles are back again acting as HDVs under the OVM model. The complete video can be found in \url{https://www.youtube.com/watch?v=YhxCZImcZL4}.}
	\label{Fig:Snapshot_ringroad}
	\vspace{-2mm}
\end{figure*}

\begin{figure*}[tb!]
	\vspace{2mm}
	\centering
	\includegraphics[width=12cm]{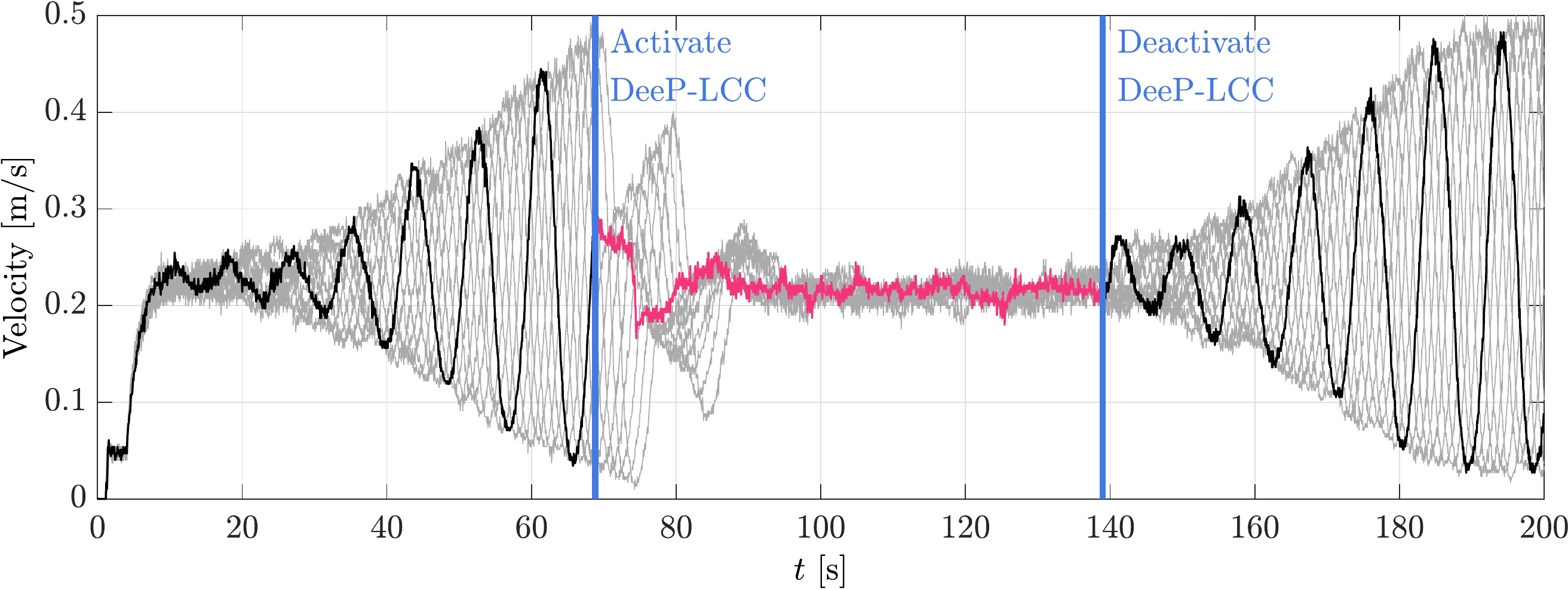}
	\vspace{-2mm}
	\caption{Velocity profiles of all the vehicles in the ring-road experiments. The black profile and the red profile represent the velocity of vehicle $5$ as an HDV or CAV with \method{DeeP-LCC}, respectively. The gray profiles represent the velocity of the other vehicles in traffic flow. \method{DeeP-LCC} is activated for vehicle $5$ at time $68.85\,\mathrm{s}$ and deactivated at time $139.05\,\mathrm{s}$. }
	\label{Fig:Velocity_ringroad}
	\vspace{-2mm}
\end{figure*}

\begin{remark}[Wave-damping CAV control strategies emerging from \method{DeeP-LCC}]
From the snapshots in Fig.~\ref{Fig:Snapshot_001000}, we can see an interesting response behavior of vehicle $2$ when facing a traffic wave: it leaves a relatively large distance from the preceding vehicle when the perturbation happens, which gives enough space to absorb the upcoming traffic wave; then, it leads the following vehicles to smoothly catch up with the traffic flow ahead, with smaller velocity overshoot with respect to the estimated equilibrium. Indeed, similar driving behavior patterns have been designed for the CAVs to mitigate traffic waves before, which require the knowledge on the future evolution of traffic waves~\cite{nishi2013theory,he2016jam} or careful tuning of controller structure and parameters~\cite{stern2018dissipation,kesting2008adaptive}. With only measurable traffic data, by contrast, our \method{DeeP-LCC} strategy achieves a desired behavior in Fig.~\ref{Fig:Snapshot_001000} (one CAV, sinusoid perturbation), and further demonstrates a cooperation potential at multi-CAV cases with more complex behaviors (see Fig.~\ref{Fig:Velocity_010100} and Fig.~\ref{Fig:Velocity_001010}) and an applicability to various traffic scenarios (see the ring-road experiments next).
\end{remark}

\subsection{Ring-road Experiments}

As clarified in Section~\ref{Sec:ExperimentDesign}, the ring-road experiment is motivated by the setup in~\cite{stern2018dissipation}. At the beginning, all the vehicles are initially HDVs under the OVM model. They are distributed uniformly in the ring-road system, and start to move from an idle velocity of $0.05\,\mathrm{m/s}$ (see Fig.~\ref{Fig:Snapshot_ringroad_initial}). As observed in Fig.~\ref{Fig:Velocity_ringroad} (the velocity profiles), a traffic wave gradually appears: the velocity of each vehicle begins to oscillate and the amplitude grows up during the propagation. Finally, all the vehicles are involved in a stop-and-go pattern. The snapshot at time $66.15 \, \mathrm{s}$ in Fig.~\ref{Fig:Snapshot_ringroad_hdv} clearly shows that in the ring-road system, some vehicles are clustered together with almost zero velocities and small inter-vehicle distances, while some other vehicles are moving with large velocities to try to catch up with its predecessor. This experiment reproduces the phenomenon of stop-and-go waves with no bottlenecks, similar to the field experiments in~\cite{sugiyama2008traffic,stern2018dissipation}.

After the stop-and-go wave emerges, vehicle $5$ is switched to CAV under the \method{DeeP-LCC} control strategy at time $68.85\,\mathrm{s}$. We can clearly observe from the velocity profiles in Fig.~\ref{Fig:Velocity_ringroad} that the stop-and-go wave is rapidly mitigated. After \method{DeeP-LCC} is activated, vehicle $5$ completely dampens the traffic wave and steers the ring-road traffic system to the equilibrium, where all the vehicles are moving with a homogeneous velocity and a stable spacing; see Fig.~\ref{Fig:Snapshot_ringroad_cav} for the snapshot at time $134.75\,\mathrm{s}$. Although some small velocity oscillations still persist, \method{DeeP-LCC} keeps the traffic flow in stability and prevents any traffic waves to reoccur. From time $139.05\,\mathrm{s}$, \method{DeeP-LCC} is deactivated, and all the vehicles are again under OVM control. Then, a stop-and-go wave reappears after a while of system evolution without the influence of \method{DeeP-LCC}, as shown in Fig.~\ref{Fig:Snapshot_ringroad_hdv2} and Fig.~\ref{Fig:Velocity_ringroad}. This experiment demonstrates the wave-dampening effect of \method{DeeP-LCC} in closed traffic scenarios without bottlenecks. 

\subsection{Discussions on Practical Factors}

In our experiments, several practical factors that are common in real-world applications of CAV technologies have been naturally incorporated. In the following, we present further discussions on low-level dynamics, data noise and system delay.

\subsubsection{Low-level dynamics} As clarified in Section~\ref{Sec:Platform}, the cloud server translates the command acceleration to command velocity via~\eqref{Eq:CommandVelocity}, and sends it to each vehicle for low-level execution. Thus, the real driving behavior of each vehicle is different from the ideal control model in the cloud server; see Fig.~\ref{Fig:OVM_trajectory} for example. 
    Most existing model-based control strategies need to estimate this dynamics from upper-level command to low-level behavior for controller design~\cite{milanes2013cooperative,jin2017optimal,jin2018experimental,naus2010string}. However, \method{DeeP-LCC} has no need to know the low-level vehicle dynamics. It directly utilizes available data, including input from cloud and output from traffic, to capture the mixed traffic behavior and design valid control inputs for the CAVs. Indeed, the control input~\eqref{Eq:ControlInput} in \method{DeeP-LCC} is not necessarily defined as the acceleration of the CAVs; it could be any command signal transmitted from the cloud to the CAVs, as long as Assumptions~\ref{Assumption:PersistentExcitation} and~\ref{Assumption:Controllable} are satisfied. System identifications for vehicle dynamics and HDVs' behavior are both bypassed in \method{DeeP-LCC}.

\subsubsection{Data noise} This has been offline tested and illustrated in Table~\ref{Tb:Noise_Error} for our experiment platform. Existing validations for data-driven control in mixed traffic are mostly based on numerical simulations where artificially designed noise is added into the system; see, \eg,~\cite{wu2021flow,huang2020learning,wang2022data,lan2021data}. In particular, for the ring-road traffic system, the simulated noise in the control process plays a significant role in reproducing the traffic wave phenomenon without bottlenecks. By contrast, our experiment platform naturally incorporates the effect of noise, and the potential of \method{DeeP-LCC} has been revealed in addressing real-world noisy measurements.

\begin{figure}[t]
	\centering
	\subfigure[System delay with $S=\{2\}$]
	{
	\includegraphics[width=4cm]{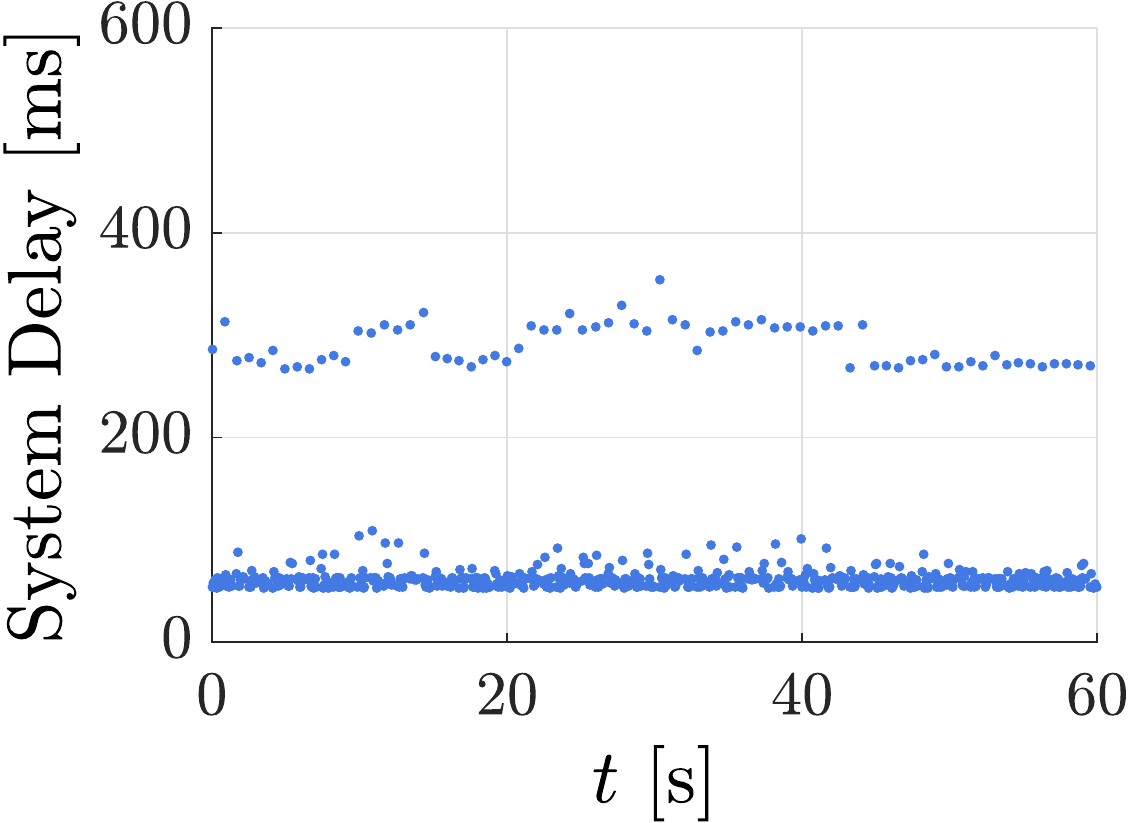}
	}
	\subfigure[System delay with $S=\{2,4\}$]
	{
	\includegraphics[width=4cm]{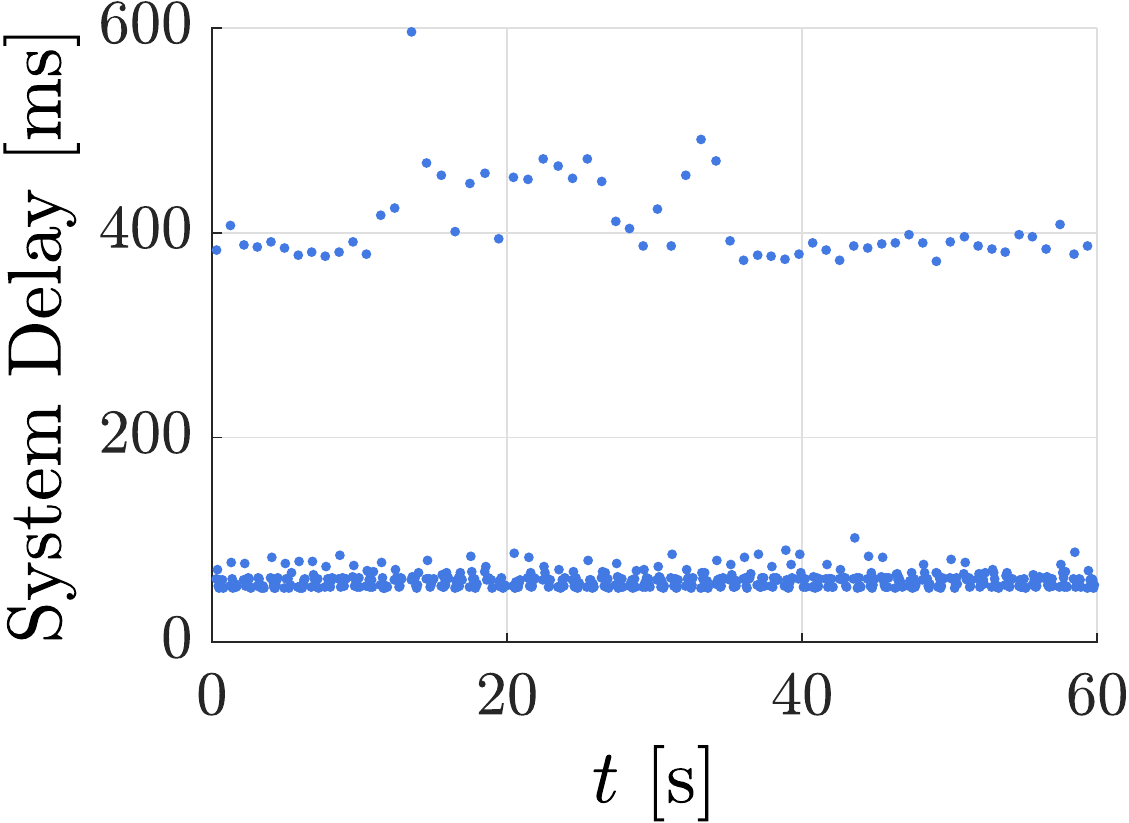}
	\label{Fig:OVM_spacing}
	}
	\vspace{-2mm}
	\caption{Total system delay in the straight-road experiments with respect to each time step.  }
	\label{Fig:TotalDelay}
	\vspace{3mm}
\end{figure}
	
\begin{table}[tb]
	\begin{center}
	\caption{Average Total System Delay in Straight-Road Experiments}
\vspace{-3mm}
		\label{Tb:TotalDelay}
		\setlength{\tabcolsep}{2mm}{
		\begin{tabular}{ccc}
		\toprule
			& \shortstack{Delay when \\ solving \method{DeeP-LCC}}   & \shortstack{Delay when \\ applying control inputs} \\\hline
			 $S=\{2\}$ & $292.80\,\mathrm{ms}$ & $60.88\,\mathrm{ms}$ \\
			$S=\{2,4\}$ & $405.69\,\mathrm{ms}$ & $60.41\,\mathrm{ms}$  \\
			\bottomrule
		\end{tabular}
		}
	\end{center}
	\vspace{-3mm}
\end{table}

\subsubsection{System delay} The localization and communication delays have been independently tested, as shown in Table~\ref{Tb:Delay}. Here we further show the total system delays in the experiments. We take the straight-road experiments with $S=\{2\}$ or $S=\{2,4\}$ as examples, and show the real-time system delay at each time step in Fig.~\ref{Fig:TotalDelay}. The average total delays are illustrated in Table~\ref{Tb:TotalDelay}. Recall that the \method{DeeP-LCC} optimization problem~\eqref{Eq:DeeP-LCC_Experiment} is solved every $N_\mathrm{c}$ time steps via a receding horizon manner (see Algorithm~\ref{Alg:DeeP-LCC}). It can be clearly observed that when \method{DeeP-LCC} is being solved, the time delay is much higher than other time steps due to the process of numerical computations. Moreover, with more CAVs incorporated, as shown in the comparison between the cases of $S=\{2\}$ and $S=\{2,4\}$, 
 solving the optimization problem~\eqref{Eq:DeeP-LCC_Experiment} costs more computation time. Nevertheless, \method{DeeP-LCC} has shown practical robustness against the influence of time delays in our experiments.


\balance

\section{Conclusions}
\label{Sec:6}

In this paper, we have presented the methodology for practical implementation of \method{DeeP-LCC}, and experimentally validated the performance of \method{DeeP-LCC} for CAV control in mixed traffic. Our results have shown that by directly utilizing real-world traffic data, \method{DeeP-LCC} enables the CAVs to mitigate traffic waves and smooth traffic flow, either on the straight-road scenario or on the ring-road scenario. From a miniature experimental perspective, these results suggest the great potential of data-driven techniques, particularly \method{DeeP-LCC}, in stabilizing traffic flow. 

All the HDVs in our experiments have a human-like car-following model, and this is common in many studies via numerical validations~\cite{zheng2020smoothing,jin2017optimal,wu2021flow} or miniature experiments~\cite{beaver2020demonstration,jang2019simulation}. It is an interesting future direction to test the performance of \method{DeeP-LCC} with real human data by involving human drivers to control the HDVs. Another interesting direction is to develop adaptive parameter tuning techniques for \method{DeeP-LCC} at various cases of mixed traffic, especially the penetration rates and spatial locations of CAVs. 
Robust design for \method{DeeP-LCC} against measurement noise and system delays is worth further investigations as well. Finally, considering the real-time requirement in large-scale implementation, it is also interesting to develop techniques to reduce the computation time of solving the optimization problem at each time step for \method{DeeP-LCC}.



\ifCLASSOPTIONcaptionsoff
  \newpage
\fi



%

\bibliographystyle{IEEEtran}
\bibliography{IEEEabrv,mybibfile}

\end{document}